\documentclass[11pt]{article}

\usepackage{graphicx}      
\usepackage{subfig}
\usepackage{color}
\usepackage{amsmath}
\usepackage{amsfonts}
\usepackage{amssymb}
\usepackage{amsthm}
\newenvironment{pf}{{\bfseries Proof}}{\qed}

\usepackage{bbm}
\usepackage{enumerate}
\usepackage[ruled, vlined, linesnumbered]{algorithm2e}
\let\oldnl\nl
\newcommand{\nonl}{\renewcommand{\nl}{\let\nl\oldnl}} 	

\usepackage{authblk} 

\newcommand{\bv}[1]{\mathbf{#1}}		
\newcommand{\bvltn}[1]{\boldsymbol{#1}}	

\newcommand{\norm}[1]{\|{#1} \|}
\newcommand{\normbig}[1]{\left\|{#1} \right\|}
\newcommand{\curlybracketsbig}[1]{\left\{ #1 \right\}}
\newcommand{\curlybrackets}[1]{\{ #1 \}}
\newcommand{\squarebracketsbig}[1]{\left[ #1 \right]}
\newcommand{\squarebrackets}[1]{[ #1 ]}
\newcommand{\parenthesesbig}[1]{\left( #1 \right)}
\newcommand{\parentheses}[1]{( #1 )}

\newcommand{\wht}[1]{\widehat{#1}}		
\newcommand{\wtd}[1]{\widetilde{#1}}	

\newcommand{\R}{\mathbb{R}}				
\newcommand{\N}{\mathcal{N}}			
\newcommand{\indicator}{\mathbbm{1}}	
\newcommand{\onevec}{\mathbf{1}} 		

\newcommand{\diag}{\textbf{diag}} 		
\newcommand{\expctn}{\mathbb{E}} 		

\makeatletter
\newcommand*{\T}{%
	{\mathpalette\@transpose{}}%
}
\newcommand*{\@transpose}[2]{%
	\raisebox{\depth}{$\m@th#1\intercal$}%
}
\makeatother

\newcommand{\irchi}[2]{\raisebox{\depth}{$#1\chi$}}
\DeclareRobustCommand{\rchi}{{\mathpalette\irchi\relax}} 	

\newcommand*{\x}{\mathsf{x}\mskip1mu} 	

\newcommand{\splitatcommas}[1]{%
	\begingroup
	\begingroup\lccode`~=`, \lowercase{\endgroup
		\edef~{\mathchar\the\mathcode`, \penalty0 \noexpand\hspace{0pt plus .1em}}%
	}\mathcode`,="8000 #1%
	\endgroup
} 

\newcommand\Item[1][]{%
	\ifx\relax#1\relax  \item \else \item[#1] \fi
	\abovedisplayskip=0pt\abovedisplayshortskip=0pt~\vspace*{-\baselineskip}
} 

\providecommand{\keywords}[1]{\textbf{\textit{Keywords:}} #1}

\newtheorem{thm}{Theorem}
\newtheorem{lem}[thm]{Lemma}
\newtheorem{cor}[thm]{Corollary}
\newtheorem{assum}[thm]{Assumption}

\newcommand{\singval}{s} 		
\newcommand{\sysidx}{z}			
\newcommand{\nsys}{m}

\title{\textbf{A Robust Algorithm for Online Switched System Identification} \footnote{This work is supported by DARPA grant N66001-14-1-4045, DARPA grant 16-43-D3M-FP-037 and NSF Grant ECCS-1508943.}}
\author[ ]{Zhe Du}
\author[ ]{Necmiye Ozay}
\author[ ]{Laura Balzano}

\affil[ ]{Electrical and Computer Engineering, University of Michigan}
\affil[ ]{\textit {\{zhedu,necmiye,girasole\}@umich.edu}}

\date{}
\begin{document}
\maketitle

\begin{abstract}                
	In this paper, we consider the problem of online identification of Switched AutoRegressive eXogenous (SARX) systems, where the goal is to estimate the parameters of each subsystem and identify the switching sequence as data are obtained in a streaming fashion. Previous works in this area are sensitive to initialization and lack theoretical guarantees. We overcome these drawbacks with our two-step algorithm: (i) every time we receive new data, we first assign this data to one candidate subsystem based on a novel robust criterion that incorporates both the residual error and an upper bound of subsystem estimation error, and (ii) we use a randomized algorithm to update the parameter estimate of chosen candidate. We provide a theoretical guarantee on the local convergence of our algorithm. Though our theory only guarantees convergence with a good initialization, simulation results show that even with random initialization, our algorithm still has excellent performance. Finally, we show, through simulations, that our algorithm outperforms existing methods and exhibits robust performance. \\	
\end{abstract}

	
	\section{Introduction}
	A SARX system is a special type of hybrid system composed of multiple subsystems/modes each with different parameters. At each time step only one subsystem is dominating and the dominant subsystem may switch over time. Given system inputs and outputs at each time step, our goal is to identify the switching sequence (discrete states) as well as to estimate the parameters of the subsystems every time we receive new data. This is a problem involving both clustering and estimation. 	
	
	In additional to applications in adaptive control, SARX system identification has been applied to video and texture segmentation \cite{vidal_recursive_2008,ozay_set_2015, ozay_sparsification_2012}. Due to the autoregressive nature of SARX model, it can also be applied to earthquake record analysis \cite{kozin1988autoregressive}, brain electrical activity mapping \cite{ogawa1993application}, meteorological objects identification \cite{bezruck2010application}, and financial time series analysis \cite{cochrane2005time}.
	
	
	
	\subsection{Prior Work}
	
	There have been many studies on the switched system identification problem in the offline/batch setting. A type of algebraic method was proposed in \cite{vidal_algebraic_2003}, which uses Veronese embedding to decouple the task of estimating the system parameters and switching sequence, and an exact solution is provided when the process and data are noise-free. Furthermore, the case when system orders are not necessarily equal or known is discussed in \cite{ma_identification_2005}. For systems with noise and measurements corrupted by outliers, \cite{ozay_set_2015} extends the algebraic method  by converting it to a rank minimization problem that is relaxed to a semi-definite program. Methods utilizing sparsity are proposed in \cite{bako_identification_2011, ozay_sparsification_2012}. 
	
	As opposed to the offline/batch setting, where we have access to all the data at once, there are many problems in which the data appears in a streaming (online) fashion. That is, at each time step, we receive data with which we need to identify current dominant subsystem as well as give the latest estimate of the system parameters. Note that naively employing an offline algorithm by using all the data in the past at each step would be computationally intractable. The majority of online algorithms use a two-step approach that alternates between determining the switching sequence and updating the parameter estimates. The work in \cite{vidal_recursive_2008} is one of the first to study online identification of switched systems using an extension of the offline algebraic method \cite{vidal_algebraic_2003}. In the algorithms proposed in \cite{bako_recursive_2011}, \cite{goudjil2016convergence}, candidate estimates are built for each of the subsystems first. Then, every time a new data point arrives, the discrete state is determined by assigning the data to one of the candidates according to some criterion, and then the estimate of chosen candidate is updated with the new data. The algorithm in \cite{bako_recursive_2011} first identifies the discrete states based on prior or posterior residual error, and then updates the estimate using recursive least squares. The algorithm in \cite{goudjil2016convergence} identifies the discrete states by minimizing prior residual error similarly and then update the estimates with a modified Outer Bounding Ellipsoid (OBE) algorithm. 
	
	\subsection{Contributions and Outline}
	
	We observe that in two-step algorithms, choosing a candidate based on minimum residual error can be sensitive to candidate initializations, since when a new subsystem dominates, it might ``take-over" a partially convergent candidate estimate if there is no candidate yet closer to its true parameters.	

	The main contribution of our paper is a more robust two-step algorithm that can effectively overcome this issue. We initialize candidate estimates for each of the subsystems. Every time we receive new data, we determine the discrete state by assigning this data to one of candidates based on a robust criterion that incorporates both residual error and an upper bound of estimation error. After we assign the data to a candidate, we update the selected candidate using a variant of the randomized Kaczmarz algorithm proposed in \cite{strohmer2009randomized} or normalized least mean squares (NLMS) \cite{simon_s._haykin_[simon_2003}. We provide  partial and local convergence results for our algorithm. In our  partial convergence analysis we assume that we can always make correct assignments, i.e. identify the discrete state correctly, thus the parameter estimation updated for the candidates can be treated as if we are using data from a single subsystem. In local convergence analysis, we assume all candidate estimates have ``good enough" initializations, and show that with some probability no misassignment will ever be made and prove the convergence of parameter estimates. Our numerical simulations verify the convergence result and show obvious improvements of our algorithm over state of the art.
	
	The paper is organized as follows: in Section \ref{sctn_ProblemFormulation}, we present the problem formulation of online SARX system identification; Section \ref{sctn_Drawback} briefly discusses the drawbacks of existing algorithms; Section \ref{sctn_OurAlgorithm} introduces our algorithm; Section \ref{sctn_OurAlgorithm} gives the theoretical analyses of our algorithm; some discussions and extensions are provided in Section \ref{sctn_Extensions}; simulation evaluation are given in Section \ref{sctn_Experiments}.

	
	\section{Problem Formulation} \label{sctn_ProblemFormulation}
	
	\subsection{SARX System}
	
	A SARX system is defined by the following expression:
	\begin{equation} \label{eqn_ARXSystem}
	y_t = \sum\nolimits_{j=1}^{n_a}a_j(\sysidx_t) y_{t-j} + \sum\nolimits_{k=1}^{n_c} c_k(\sysidx_t) u_{t-k} + n_t
	\end{equation}
	where $u_t \in \R $ and $y_t \in \R$ are the input and output of the system, and $n_t \in \R$ is an additive noise term. The discrete state $\sysidx_t \in \{1, \dots, \nsys\} \equiv [\nsys]$ indexes the dominant/active subsystem at time $t$, and $\{\sysidx_t \}_t$ denotes the switching sequence. Coefficients $\{a_j(\sysidx_t)\}_{j=1}^{n_a}$ and $\{c_j(\sysidx_t)\}_{j=1}^{n_c}$ are the parameters of subsystem $\sysidx_t$.  Let $\bvltn{\phi}_{y,t} = [y_{t-1}, \dots, y_{t-n_a}]^\T, \bvltn{\phi}_{u,t} = [u_{t-1}, \dots, u_{t-n_c}]^\T, \bvltn{\phi}_t = [\bvltn{\phi}_{y,t}^\T, \bvltn{\phi}_{u,t}^\T]^\T$, and furthermore, let $\splitatcommas{\bv{w}_{\sysidx_t} = [a_1(\sysidx_t), \dots, a_{n_a}(\sysidx_t), c_1(\sysidx_t), \dots, c_{n_c}(\sysidx_t)]^\T}$. With this notation, the SARX system dynamics \eqref{eqn_ARXSystem} can be written in vector form: 
	\begin{equation} \label{eqn_VecARXSystem}
	y_t = \bv{w}_{\sysidx_t}^\T \bvltn{\phi}_t + n_t \;.
	\end{equation}
	Let $n = n_a + n_c$ be the system order, which can also be viewed as the ambient dimension of our problem.
	
	\subsection{Assumptions}
	
	In this work, we make the following assumptions, where Assumption \ref{asmp_systemSpecs} and the noise upper bound in Assumption \ref{asmp_noise} are needed for the algorithm to work. (Case where noise is unbounded is discussed in Section \ref{subsctn_unboundedNoise}.) Assumption \ref{asmp_noise} to Assumption \ref{asmp_NoAmbiguous} are mainly for analysis purposes.
	\begin{assum} \label{asmp_systemSpecs}
		The model orders $n_a$, $n_c$ on RHS of \eqref{eqn_ARXSystem}, and the number of subsystems, $\nsys$, are known.
	\end{assum}
	
	\begin{assum} \label{asmp_noise}
		The noise $n_t$ is random with $\expctn[n_t]=0$ and $\expctn[n_t^2]=\sigma_n^2$. $ |n_t| \leq n_{\max}$ and $n_{\max}$ is known. $n_t$ is independent of input $u_t$.
	\end{assum}
	
	\begin{assum} \label{asmp_data}
		For all $t$, $\norm{\bvltn{\phi}_t} \leq \phi_{\max}$. We also assume a lower bound on the SNR: for all $t, \frac{\norm{\bvltn{\phi}_t}}{ | n_t|} \geq S_{\min} $.
	\end{assum}

	\begin{assum} \label{asmp_singularvalue}
		There exists $\singval_{\max} \geq \singval_{\min}> 0$ such that $\forall \ S \subset \mathbb{N}^+$ with cardinality $N_R$ (defined in Section \ref{sctn_OurAlgorithm}),
		\begin{equation} \label{eqn_untitle52}
		\singval_{\min}^2 I_n \preceq \sum\nolimits_{t\in S} \bvltn{\phi}_t \bvltn{\phi}_t^\T \preceq \singval_{\max}^2 I_n \;.
		\end{equation}
	\end{assum}
	
	\begin{assum} \label{asmp_NoAmbiguous}
		If subsystem $i$ generates data pair $\{\bvltn{\phi}_t, y_t \}$, i.e. $y_t = \bv{w}_i^\T \ \bvltn{\phi}_t + n_t$, then $\forall j \neq i$, $|\bv{w}_j^\T \bvltn{\phi}_t - \bv{w}_i^\T \bvltn{\phi}_t | \geq \psi$.
	\end{assum}
		
	Assumption \ref{asmp_singularvalue} is similar to persistent excitation conditions in the literature, and it plays a critical role in the convergence rate. Assumption \ref{asmp_NoAmbiguous} guarantees there is no ambiguous data, since if data pair $\curlybrackets{\bvltn{\phi}_t, y_t}$ satisfies both $y_t = \bv{w}_1^\T \bvltn{\phi}_t$ and $y_t = \bv{w}_2^\T \bvltn{\phi}_t$, then even with the true parameters $\bv{w}_1$ and $\bv{w}_2$, we cannot tell which system generates $y_t$.
	
    \subsection{Goal}
    
	The goal of online system identification is as follows. After we collect the data pair $\curlybrackets{\bvltn{\phi}_t, y_t}$ at each time step, we want to identify discrete state $\sysidx_t$ and estimate parameters of the subsystem that generates $y_t$.

	
	\section{Drawbacks of Existing Algorithms} \label{sctn_Drawback}
	Existing algorithms e.g. \cite{bako_recursive_2011}, \cite{goudjil2016convergence}, commonly have a two-step structure after candidate estimate for each of the subsystem is initialized: (1) every time new data is available, it is assigned to the candidate with minimum prior/posterior residual error; (2) the parameter estimate of the chosen candidate is updated with this data. We will show that using only residual error as the criterion to assign data can be unreliable.
	\begin{figure}[h!]
		\begin{center}
			\includegraphics[width=10cm]{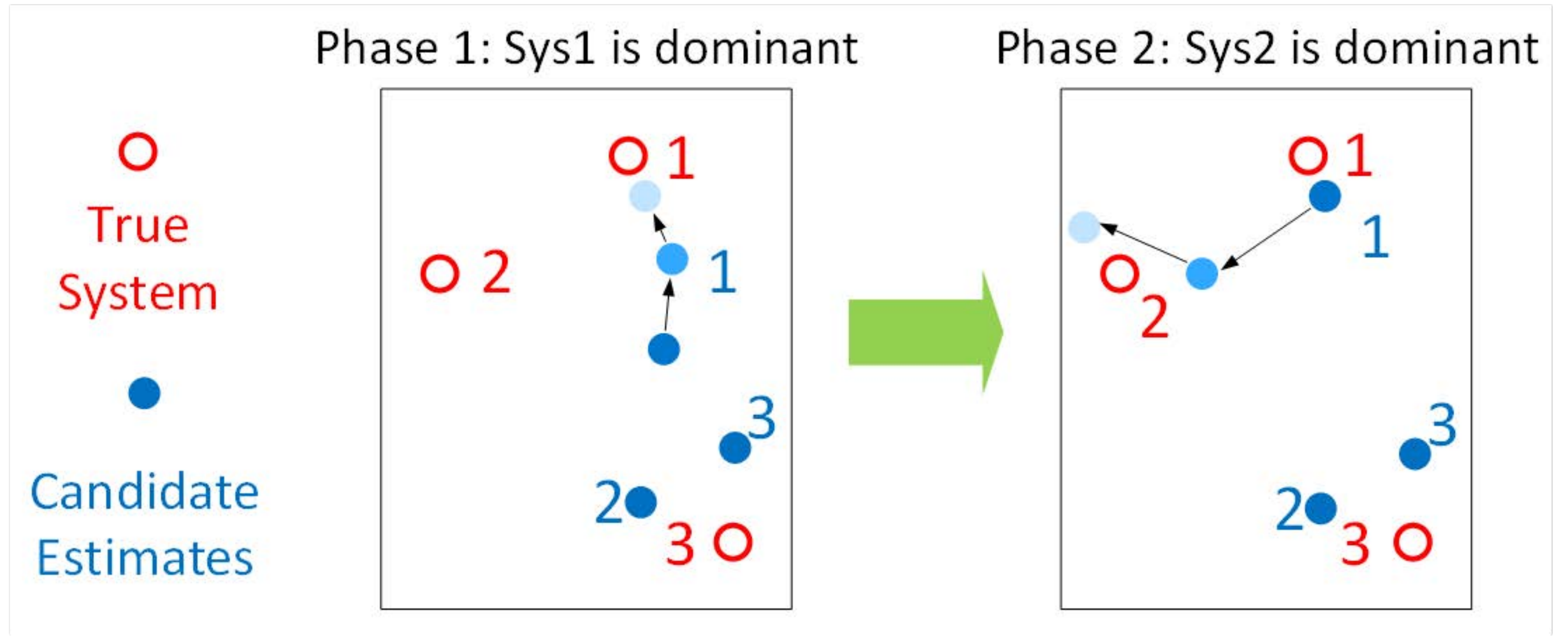}
			\caption{\small Demonstration of potential drawback of existing algorithms.} 
			\label{fig_DownsideofPreviousAlgs}
		\end{center}
	\end{figure}
	
	Fig. \ref{fig_DownsideofPreviousAlgs} shows a toy example of what could go wrong with the above mentioned algorithms. There are 3 subsystems, and red circles illustrate their true parameter vectors in the ambient space; the 3 candidate estimates are initialized at the three dark blue points in the left box. Assume from $t=1 \text{ to } 10$, subsystem 1 is dominant (left box), and from $t=11 \text{ to } 20$, subsystem 2 is dominant (right box). Considering the positions and true and estimated parameters, it's \emph{likely} that from $t=1 \text{ to } 10$ data generated by subsystem 1 will be assigned to candidate 1 since it's the closest candidate. 
    When $t=10$, candidate 1 is an improved estimate of system 1 parameters, given by the light blue point in the left box. At time $t=11$, subsystem 2 becomes dominant. Considering the current positions of all candidates, candidate 1 is still closest to subsystem 2, so it's \emph{likely} that data generated by subsystem 2 will also be assigned to candidate 1, and we could expect candidate 1 will start to drift from subsystem 1 parameter values toward subsystem 2, given by the trajectory in the right box. In this sense, all previous efforts used to let candidate 1 learn subsystem 1 will be wasted. 	
	
	In our algorithm, the basic idea to solve this drawback is to take the accuracy of the candidate estimates into account and be more cautious when assigning data to candidates with higher accuracy. The details will be discussed in Section \ref{subsctn_makeAssignment}.

	
	\section{Our Algorithm} \label{sctn_OurAlgorithm}
	\begin{algorithm}[h!]
		Initialize $N_R, N_C (N_R \geq n, N_C \geq N_R^2), \alpha, \beta, \nu$ \label{algline_untitled8}\\
		\For{$i = 1, \dots, \nsys$}{
			$ \wht{\bv{w}}_{i,0} = \bv{0}_{n {\times} 1}, c_i = 0, $
			$\bvltn{\Phi}^R_{i,t} = \bv{0}_{n {\times} N_R}, \, \bv{y}^R_{i,t} = \bv{0}_{N_R {\times} 1},$ \\
			$\bvltn{\Phi}^C_{i,t} = \bv{0}_{n {\times} N_C}, \, \wht{\bv{W}}^C_{i,t} = \bv{0}_{n {\times} N_C}, \, \bv{h}^C_{i,t} = \bv{0}_{N_C {\times} 1}, \, \epsilon_{i,0}^u = \infty$ \\
		} \label{algline_untitled9}
		\For{$t = 1, 2, \dots$}{	
			Receive $\curlybrackets{\bvltn{\phi}_t, y_t}$. \label{algline_untitled10}\\	
			
			Compute normzlized residual errors and potential new estimates for all candidates: \\			
			\Indp $r_i = |y_t - \wht{\bv{w}}_{i,t-1}^\T \bvltn{\phi}_t| \cdot \norm{\bvltn{\phi}_t}^{-1} \quad \forall i \in [\nsys]$ \label{algline_untitled1}\\ 
			
			$\wtd{\bv{w}}_{i,t} {=} \wht{\bv{w}}_{i,t-1} {-} \norm{\bvltn{\phi}_t}^{-2} \bvltn{\phi}_t \parentheses{\wht{\bv{w}}_{i,t-1}^\T  \bvltn{\phi}_t {-} y_t }
			\quad \forall i \in [\nsys]$ \label{algline_untitled2} \\ \Indm
			
			Choose a candidate to assign data: \\
			\Indp $\wht{\sysidx}_t = \arg \min_i \ r_i \cdot \max\parenthesesbig{1, \alpha  \frac{ \norm{\wtd{\bv{w}}_{i,t} - \wht{\bv{w}}_{i,t-1}}}{2 (\epsilon_{i,t-1}^u + \nu )}}^\beta$ \label{algline_makeAssignment}\\ \Indm
			
			Update counter and window varaibles: \\
			\Indp
			$c_{\wht{\sysidx}_t} = c_{\wht{\sysidx}_t} + 1$ \label{algline_untitled11}\\
			
			$\bvltn{\Phi}^R_{\hat{\sysidx}_t, t} {=} [\bvltn{\Phi}^R_{\hat{\sysidx}_t, {t-1}}[{:}, 2{:}\text{end}], \bvltn{\phi}_t], \,
				\bv{y}^R_{\hat{\sysidx}_t, t} {=} [\bv{y}^R_{\hat{\sysidx}_t, {t-1}}[2{:}\text{end}]; y_t]$ \label{algline_untitled12}\\	\Indm
			
			Update estimate of chosen candidate: \\
			\Indp
			\SetInd{2.05em}{-0.6em}
			\uIf{$c_{\wht{\sysidx}_t} < N_R$}{
				$\bvltn{\phi}_t^* = \bvltn{\phi}_t, y_t^* = y_t, \eta_t^* = \norm{\bvltn{\bvltn{\phi}}_t}^{-2}$ \label{algline_untitled4}\\				
			}
			\Else{
				Sample $l_t {\in} [N_R]$  $\text{w.p.}\norm{\bvltn{\Phi}^R_{{\wht{\sysidx}_t}, t}[:, l_t]}^2 / \norm{\bvltn{\Phi}^R_{{\wht{\sysidx}_t}, t}}_F^2$ \label{algline_PickColumn}\\
				$\bvltn{\phi}_t^* {=} \bvltn{\Phi}^R_{{\wht{\sysidx}_t}, t}[:, l_t], \,  y_t^* {=} \bv{y}^R_{{\wht{\sysidx}_t}, t}[l_t], \, \eta_t^* {=} \norm{\bvltn{\phi}_t^{*}}^{-2}$ \label{algline_untitled5}\\		
			}
			$\wht{\bv{w}}_{{\wht{\sysidx}_t},t} = \wht{\bv{w}}_{{\wht{\sysidx}_t},t-1} - \eta_t^* \bvltn{\phi}_t^* \parentheses{\wht{\bv{w}}_{{\wht{\sysidx}_t},t-1}^\T  \bvltn{\phi}_t^* - y_t^* }$ \label{algline_updateEstimate}\\
			\Indm
			
			Update error upper bound and window variables: \\
			\Indp
			$\bvltn{\Phi}^C_{\wht{\sysidx}_t,t}, \wht{\bv{W}}^C_{\wht{\sysidx}_t,t}, \bv{h}^C_{\wht{\sysidx}_t,t}, \epsilon_{\wht{\sysidx}_t, t}^u {=} \textbf{\small {UpdateUpperBound}}$ \label{algline_computeUpperbound} \\

			$\forall i\neq {\wht{\sysidx}_t}, \curlybrackets{\wht{\bv{w}}_{i,t}, \bvltn{\Phi}^R_{i,t}, \bv{y}^R_{i,t}, \bvltn{\Phi}^C_{i,t}, \wht{\bv{W}}^C_{i,t}, \bv{h}^C_{i,t}, \epsilon_{i,t}^u} = \curlybrackets{\wht{\bv{w}}_{i,\tau}, \bvltn{\Phi}^R_{i,\tau}, \bv{y}^R_{i,\tau}, \bvltn{\Phi}^C_{i,\tau}, \wht{\bv{W}}^C_{i,\tau}, \bv{h}^C_{i,\tau}, \epsilon_{i,\tau}^u}_{\tau=t-1}$ \label{algline_untitled3}\\
			\Indm	
		}
		\caption{Our Main Algorithm} \label{alg_main}
	\end{algorithm}
	
	In this paper we propose Algorithm \ref{alg_main} for online identification of SARX models. This is also a two-step algorithm, but with an improved data assignment to consider not only the residual but also system estimation accuracy. This section gives an overview of the algorithm steps.
    
    Lines \ref{algline_untitled8} to \ref{algline_untitled9} show initialization. $\wht{\bv{w}}_{i,0}$ is the initial estimate for candidate $i$, and $c_i$ is number of assignments to candidate $i$. $\splitatcommas{\bvltn{\Phi}^R_{i,t} {\in} \R^{n \x N_R}, \bv{y}^R_{i,t}{\in} \R^{N_R}, \bvltn{\Phi}^C_{i,t}{\in} \R^{n \x N_C}, \wht{\bv{W}}^C_{i,t} {\in} \R^{n \x N_C}, \bv{h}^C_{i,t} {\in} \R^{N_C}, \epsilon_{i,t}^u }$ are the corresponding window variables for candidate $i$ at time $t$, which will be explained in details later in this section. $N_R $ and $N_C$ are the number of columns of $\bvltn{\Phi}^R_{i,t}$ and $\bvltn{\Phi}^C_{i,t}$ respectively, which are also the window lengths for the randomized Kaczmarz algorithm and error upper bound estimation respectively. 
	
	At each time step, via Lines \ref{algline_untitled10} to \ref{algline_untitled11}, we assign the data to one of the candidates using a new criterion to determine the discrete state. Then, we update the chosen candidate estimate using an idea similar to the randomized Kaczmarz algorithm in \cite{strohmer2009randomized} in Lines \ref{algline_untitled12} to Line \ref{algline_untitled3}.
	
	\subsection{Making Assignment/Identifying the Discrete State} \label{subsctn_makeAssignment}
	
	With data pair $\curlybrackets{\bvltn{\phi}_t, y_t}$, we compute the normalized residual error $r_i$ for each candidate in Line \ref{algline_untitled1}, where $\wht{\bv{w}}_{i, t-1}$ is the estimate of candidate $i$ at time $t-1$. We then compute the potential new estimate $\wtd{\bv{w}}_{i,t}$ for each candidate if we were to use $\curlybrackets{\bvltn{\phi}_t, y_t}$ to update $\wht{\bv{w}}_{i, t-1}$. 
	
	The assignment criterion is given in Line \ref{algline_makeAssignment}. The criterion has two components: the first term is the normalized residual error $r_i$ and the second term measures whether $\wtd{\bv{w}}_{i,t}$ has a larger estimation error than $\wht{\bv{w}}_{i, t-1}$. The variables $\alpha$, $\beta$ and $\nu$ are tuning parameters. Variable $\epsilon_{i,t-1}^u$ is an estimate of upper bound on the magnitude of candidate $i$'s estimation error $\bvltn{\epsilon}_{i,t-1}\equiv\bv{w} - \wht{\bv{w}}_{i,t-1}$ with respect to \emph{some} true system parameter $\bv{w}$. The main difference between our algorithm and previous two-step algorithms mentioned in Section \ref{sctn_Drawback} is the incorporation of the second term, which makes assignment more robust.
	
	\begin{figure}[h!]
		\begin{center}
			\includegraphics[width=10cm]{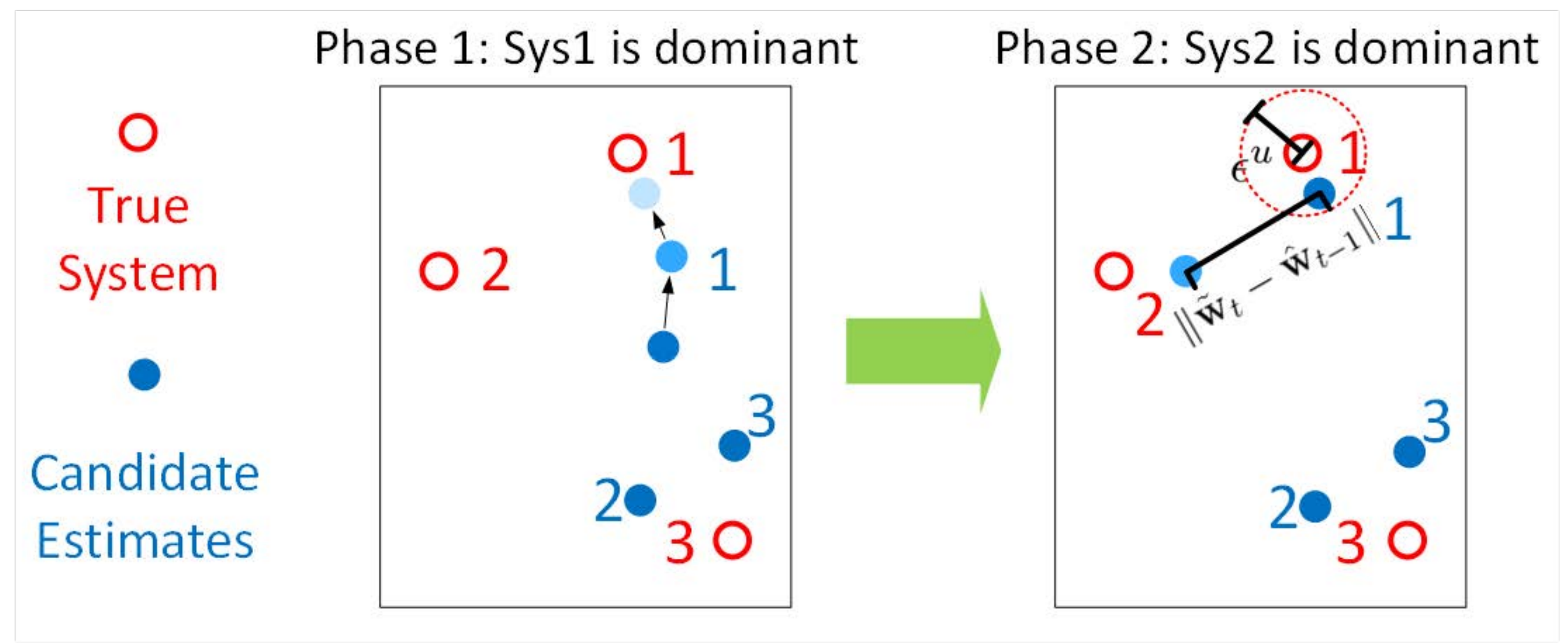}
			\caption{\small Idea of our algorithm} 
			\label{fig_IdeaofOurAlg}
		\end{center}
	\end{figure}	
	
	The idea behind the criterion is straightforward: letting $\alpha{=}1, \beta{=}1, \nu {=} 0$, and replacing $\epsilon_{i,t-1}^u$ with $\norm{\bvltn{\epsilon}_{i,t-1}}$, the second part of the criterion becomes $\max\parenthesesbig{1, \frac{\norm{\wtd{\bv{w}}_{i,t} - \wht{\bv{w}}_{i,t-1}}}{2 \norm{\bvltn{\epsilon}_{i,t-1}}}}$. The numerator $\norm{\wtd{\bv{w}}_{i,t} - \wht{\bv{w}}_{i,t-1}}$ is the magnitude of variation if we update the estimate $\wht{\bv{w}}_{i, t-1}$ of candidate $i$ with $\curlybrackets{\bvltn{\phi}_t, y_t}$. We could see that if the estimation error of $\wtd{\bv{w}}_{i,t}$ doesn't increase compared with the error of $\wht{\bv{w}}_{i,t-1}$, then we must have $\norm{\wtd{\bv{w}}_{i,t} - \wht{\bv{w}}_{i,t-1}} \leq 2 \norm{\bvltn{\epsilon}_{i,t-1}}$. And if $\norm{\wtd{\bv{w}}_{i,t} - \wht{\bv{w}}_{i,t-1}} > 2 \norm{\bvltn{\epsilon}_{i,t-1}}$, then the estimation error must get larger. Therefore, $\max\parenthesesbig{1, \frac{\norm{\wtd{\bv{w}}_{i,t} - \wht{\bv{w}}_{i,t-1}}}{2 \norm{\bvltn{\epsilon}_{i,t-1}}}}$ works against candidates whose error would increase if we update using $\curlybrackets{\bvltn{\phi}_t, y_t}$. The max operator shows that as long as $\norm{\wtd{\bv{w}}_{i,t} - \wht{\bv{w}}_{i,t-1}} \leq 2 \norm{\bvltn{\epsilon}_{i,t-1}}$, we don't penalize any further. Since we don't know the true estimation error $\norm{\bvltn{\epsilon}_{i,t-1}}$, we replace it with its estimated upper bound $\epsilon_{i,t-1}^u$ (we will show under some conditions, this is a valid upper bound in Theorem \ref{thrm_validUpperbound}) computed in Algorithm \ref{alg_computeUpperbound}. 
	
	This idea is illustrated with Fig. \ref{fig_IdeaofOurAlg}. Consider the same experimental setup as the toy example in Section \ref{sctn_Drawback}. At time $t=11$, candidate 1 will be within the ball region (denoted with the red dotted circle) estimated by upper bound $\epsilon^u$. Once the potential update magnitude of candidate 1 exceeds the diameter $2\epsilon^u$ of this region, which implies its estimation accuracy will become worse if we update candidate 1 with this data, so candidate 1 should be penalized when making the assignment.

	We note that $\norm{\wtd{\bv{w}}_{i,t} - \wht{\bv{w}}_{i,t-1}} {\leq} 2 \norm{\bvltn{\epsilon}_{i,t-1}}$ is only a necessary but not sufficient condition to ensure non-increasing estimation error. Even though we are relying on a necessary condition, our algorithm empirically achieves significantly improved performance over previous algorithms.
	
	\subsection{Candidate Estimate Updates}
	
	After the assignment is made, from Line \ref{algline_untitled12} to Line \ref{algline_updateEstimate} we update the estimate of the candidate $\wht{\sysidx}_t$ to which the data has been assigned. Our updating approach is based on the randomized Kaczmarz method with a sliding window of data. We define window variables $\bvltn{\Phi}^R_{i,t}, \bv{y}^R_{i,t}$ to store previous $N_R$ data $\curlybrackets{\bvltn{\phi}, y}$ assigned to candidate $i$. If we have collected $N_R$ data, i.e. $c_{\wht{\sysidx}_t} \geq N_R$, we update the estimate with randomly picked historical data $\curlybrackets{\bvltn{\phi}^*_t, y^*_t}$; otherwise we simply update using current data $\curlybrackets{\bvltn{\phi}^*_t, y^*_t}=\curlybrackets{\bvltn{\phi}_t, y_t}$. The idea behind the update rule is that we project the current estimate $\wht{\bv{w}}_{{\wht{\sysidx}_t},t-1}$ onto the solution space of $\curlybrackets{\bvltn{\phi}^*_t, y^*_t}$ such that $y^*_t = \wht{\bv{w}}_{{\wht{\sysidx}_t},t}^\T \bvltn{\phi}^*_t$.
	
	\subsection{Computation of Error Upper Bound}
	
	\begin{algorithm}[h!]
		Update window variables for chosen candidate: \\	\Indp
		$\bvltn{\Phi}^C_{\wht{\sysidx}_t,t} = [\bvltn{\Phi}^C_{\wht{\sysidx}_t,t-1}[:\, , \, 2:\text{end}] \, , \, \bvltn{\phi}_t^*]$ \\				 
		$\wht{\bv{W}}^C_{\wht{\sysidx}_t,t} = [\wht{\bv{W}}^C_{\wht{\sysidx}_t,t-1}[:\, , \, 2:\text{end}] \, , \, \wht{\bv{w}}_{{\wht{\sysidx}_t},t}]$\\
		$\bv{h}^C_{\wht{\sysidx}_t,t} = [\bv{h}^C_{\wht{\sysidx}_t,t-1}[2:\text{end}]\, ; \, \eta_t^*]$ \\ \Indm
		
		Update the error upper bound for chosen candidate: \\	\Indp
		\SetInd{2.05em}{-0.6em}
		\uIf{$c_{\wht{\sysidx}_t} < N_C$}
		{	
			$\epsilon_{\wht{\sysidx}_t, t}^u = \epsilon_{\wht{\sysidx}_t, t-1}^u$
		}
		\Else
		{			
			$\Delta \wht{\bv{W}} = \wht{\bv{w}}_{{\wht{\sysidx}_t},t} \onevec_{N_C \times 1}^\T - \wht{\bv{W}}^C_{\wht{\sysidx}_t,t}$ \label{algline_untitled6}\\
			$\Delta \wht{\bv{w}} = \wht{\bv{w}}_{{\wht{\sysidx}_t},t}  - \wht{\bv{w}}_{{\wht{\sysidx}_t},t-N_C}$\\
			
			$\bv{H} = \diag (\bv{h}^C_{\wht{\sysidx}_t,t})$\\
			
			$\bv{A} = \parentheses{\bvltn{\Phi}^C_{\wht{\sysidx}_t,t} \bv{H} {\bvltn{\Phi}^C_{\wht{\sysidx}_t,t}}^\T}^{-1} \bvltn{\Phi}^C_{\wht{\sysidx}_t,t} \bv{H}$ \\
			
			$\bv{b} = \parentheses{\bvltn{\Phi}^C_{\wht{\sysidx}_t,t} \bv{H} {\bvltn{\Phi}^C_{\wht{\sysidx}_t,t}}^\T}^{-1}  \squarebracketsbig{\Delta \wht{\bv{w}} - \bvltn{\Phi}^C_{\wht{\sysidx}_t,t} \bv{H} \square \parenthesesbig{\bvltn{\Phi}^C_{\wht{\sysidx}_t,t}, \Delta \wht{\bv{W}}}}$ \label{algline_untitled7} \\
			
			Let $V = \{[\pm n_{\max}, \pm n_{\max}, \dots,  \pm n_{\max}]_{N_C}^\T \}$ \\
			$\epsilon_{\wht{\sysidx}_t, t}^u = \max_{\bv{n}\in V} \norm{A \bv{n} - \bv{b}}$ \label{algline_untitled13}
			
		}
		\Indm
		\nonl \textbf{Remark:} $\square \parentheses{A,B} $$ \equiv [ a_1^\T b_1, a_2^\T b_2, \dots, a_n^\T b_n ]^\T$, where $a_i, b_i$ are the $i$th columns of matrices $A, B$
		\caption{UpdateUpperBound} \label{alg_computeUpperbound}
	\end{algorithm}
	
	We update the error upper bound estimate $\epsilon_{\wht{\sysidx}_t, t}^u$ and related window variables $\bvltn{\Phi}^C_{\wht{\sysidx}_t,t}, \wht{\bv{W}}^C_{\wht{\sysidx}_t,t}, \bv{h}^C_{\wht{\sysidx}_t,t}$ of the chosen candidate in Line \ref{algline_computeUpperbound}. The details of this update are given in Algorithm \ref{alg_computeUpperbound}. If the window is not full, i.e. $c_{\wht{\sysidx}_t} {<} N_C$, we simply follow the previous upper bound estimate, i.e. $\infty$; otherwise, we update according to a slightly complicated rule whose justification is given in Theorem \ref{thrm_validUpperbound}.

	
	\section{Theoretical Results} \label{sctn_Theory}
	Our main theorems are Theorem \ref{thrm_PartialConstant} and Theorem \ref{thrm_LocalConvergence}, which show the partial and local convergence guarantees for the algorithm respectively. Note that the proofs for all the lemmas, theorems, and corollaries are provided at the appendices.
	
	To ease the exposition, we introduce some notation and concepts that are frequently used later:
	\begin{itemize}
		\setlength{\itemsep}{0pt}
        
        \item In Algorithm \ref{alg_main}, we sample a column index $l_t$ from the matrix $\bvltn{\Phi}^R_{i,t}$ in Line \ref{algline_PickColumn} of Algorithm \ref{alg_main}. Since $\bvltn{\Phi}^R_{i,t}$ is a matrix with columns being data vectors collected at different time, we essentially sampled a time index. Let $r_t(l_t)$ denote the true time corresponding to the collecting time of data of column $l_t$.
		
		\item Let $r(i,t)$ denote number of times subsystem $i$ is dominant up to time $t$.
		
		\item Setup(A): Assume hybrid SARX system only involves 1 subsystem, namely, subsystem $i$ with parameter $\bv{w}_i$. Then $\wht{\bv{w}}_{i,t}$ is the only candidate estimate. We let $\bvltn{\epsilon}_{i,t} = \bv{w}_i - \wht{\bv{w}}_{i,t}$ denote the estimation error.
			
	\end{itemize}
	
	\subsection{Preliminary Results} \label{subsctn_PreliminaryResult}
	
	In Section \ref{subsctn_PreliminaryResult}, we first present several lemmas that serve as the building blocks for later theorems. 
	\begin{lem} \label{lemma_boundsonNorm}
		$\forall i$, after $c_i \geq N_R$, since $\bvltn{\Phi}^R_{i,t} {\bvltn{\Phi}^R_{i,t}}^\T = \sum \bvltn{\phi} \bvltn{\phi}^\T $,  and following Assumption \ref{asmp_singularvalue}, we know the singular values of $\bvltn{\Phi}^R_{i,t}$ is upper and lower bounded by $\singval_{\max}$ and $\singval_{\min}$ selectively. Following this, the following results hold trivially
		\begin{enumerate}[(i)]
			\setlength{\itemsep}{0pt}
			\item Let $F_{\max} = \sqrt{n} \singval_{\max}, F_{\min} = \sqrt{n} \singval_{\min}$, then we have
			\begin{equation}
			F_{\min} \leq \norm{\bvltn{\Phi}^R_{i,t}}_F \leq F_{\max}
			\end{equation}
			
			\item Let $\kappa(\bvltn{\Phi}^R_{i,t}) {=} \norm{\bvltn{\Phi}^R_{i,t}}_F \norm{{\bvltn{\Phi}^R_{i,t}}^{-1}}_2, \kappa_{\max} {=} \sqrt{\parentheses{{(n{-}1) \singval_{\max}^2 {+} \singval_{\min}^2 }}/\singval_{\min}^2 }$, and $\kappa_{\min} {=} \sqrt{n}$, where $-1$ denotes the right inverse, then we have
			\begin{equation}		
			\kappa_{\min} \leq \kappa(\bvltn{\Phi}^R_{i,t}) \leq \kappa_{\max}
			\end{equation}
			
			\item Let $\xi(\bvltn{\Phi}^R_{i,t}) {=} \norm{\bvltn{\Phi}^R_{i,t}}_F / \norm{\bvltn{\Phi}^R_{i,t}}_2 $, $\xi_{\max} = \sqrt{n}$, and $\xi_{\min} = \sqrt{\parentheses{\singval_{\max}^2 + (n-1)\singval_{\min}^2 }/{\singval_{\max}^2}}$, then we have	
			\begin{equation}		
			\xi_{\min} \leq \xi(\bvltn{\Phi}^R_{i,t}) \leq \xi_{\max} 	
			\end{equation}		
		\end{enumerate}
	\end{lem}
	
	\begin{lem} \label{lemma_NoiseData}
		Following Assumption \ref{asmp_noise}, we have
		\begin{enumerate}[(i)]
			\setlength{\itemsep}{0pt}
			\item $\expctn[n_{r_t(l_t)}]  = 0, \expctn[n_{r_t(l_t)}^2]  = \sigma_n^2$ \label{lemmaresult_1}
			\item $n_{r_t(l_t)}$ and $\bvltn{\phi}_{r_t(l_t)}$ are uncorrelated \label{lemmaresult_2}
			\Item 
			\begin{equation} \label{eqn_untitle101}
			\frac{N_R}{F_{\max}^2} \sigma_n^2
			\leq
			\expctn \squarebracketsbig{\frac{n_{r_t(l_t)}^2}{\norm{\bvltn{\phi}_{r_t(l_t)}}^2}} 
			\leq  
			\frac{N_R}{F_{\min}^2} \sigma_n^2
			\end{equation} \label{lemmaresult_3}		
		\end{enumerate}	
	\end{lem}
	
	The following Lemma \ref{lemma_boundonExpectation} is extension of result in \cite{strohmer2009randomized}.
	\begin{lem} \label{lemma_boundonExpectation}
		For any random vector $\bv{z} \in \R^n$, we have
		\begin{equation} \label{eqn_untitle102}
		\kappa_{\max}^{-2} \expctn [\norm{\bv{z}}^2 ]
		\leq 
		\expctn \squarebracketsbig{\parenthesesbig{\frac{\bvltn{\phi}_{r_t(l_t)}^\T \bv{z} }{\norm{\bvltn{\phi}_{r_t(l_t)}}}  }^2} 
		\leq 
		\xi_{\min}^{-2} \expctn [ \norm{\bv{z}}^2 ]
		\end{equation}
	\end{lem}
	
	\subsection{Valid Upper Bound} \label{subsctn_ValidUpBd}
	
	The following theorem gives justification for error upper bound $\epsilon_{\wht{\sysidx}_t, t}^u$ computed from Line \ref{algline_untitled6} to Line \ref{algline_untitled13} in Algorithm \ref{alg_computeUpperbound}. However, this is a restrictive result, as it requires that the candidate $\wht{\sysidx}_t$ is updated with data from the same subsystem for last $N_C$ steps, i.e. the elements in $\bvltn{\Phi}^C_{\wht{\sysidx}_t,t}, \wht{\bv{W}}^C_{\wht{\sysidx}_t,t}, \bv{h}^C_{\wht{\sysidx}_t,t}$ are collected from the same subsystem.
	
	\begin{thm} \label{thrm_validUpperbound}
	Assume at some time $t$, $\wht{\sysidx}_t = \sysidx_t = i$, $c_i \geq N_C$, and $\bvltn{\Phi}^C_{\wht{\sysidx}_t,t}, \wht{\bv{W}}^C_{\wht{\sysidx}_t,t}, \bv{h}^C_{\wht{\sysidx}_t,t}$ are constructed from data entirely from subsystem $i$, then $\epsilon_{i,t}^u$ is a valid upper bound for $\bvltn{\epsilon}_{i,t}$, i.e. $\epsilon_{i,t}^u \geq \norm{\bvltn{\epsilon}_{i,t}} = \norm{\bv{w}_i - \wht{\bv{w}}_{i,t}}$.
	\end{thm}
	
	\subsection{Partial Convergence Results} \label{subsctn_PartialConvrg}
	
	In this section, we list the results regarding partial convergence where we assume there is no misassignment, i.e. data generated from the same subsystem can all be assigned to one particular candidate. In this sense, to analyze the convergence properties, it suffices to consider the case where there is only one subsystem in the hybrid model, and one corresponding candidate, i.e. Setup(A).
	
	Lemma \ref{lemma_InitPhasePartial} provides the convergence analysis at the beginning phase of the algorithm, when $t<N_R$ and Algorithm \ref{alg_main} executes Line \ref{algline_untitled4}. Then Lemma \ref{lemma_2ndConvergenceConstant} provides the convergence analysis for the second phase of the algorithm, when $t\geq N_R$ and Algorithm \ref{alg_main} executes Line \ref{algline_PickColumn} and \ref{algline_untitled5}. Finally, we have the partial convergence result Theorem \ref{thrm_PartialConstant} simply by combining Lemma \ref{lemma_InitPhasePartial} and Lemma \ref{lemma_2ndConvergenceConstant}.
	
	\begin{lem} \label{lemma_InitPhasePartial}
		With Setup(A), we have
		\begin{equation} \label{eqn_InitPhasePartial}
		\frac{\sigma_n^2}{\phi_{\max}^2}
		\leq 
		\expctn \squarebracketsbig{\norm{\bvltn{\epsilon}_{i, N_R-1}}^2} 
		\leq
		\norm{\bvltn{\epsilon}_{i,0}}^2 + \frac{N_R-1}{S_{\min}^2}		
		\end{equation}
	\end{lem}
	
	\begin{lem} \label{lemma_2ndConvergenceConstant}
		With Setup(A), for $t\geq N_R$ we have
		\begin{equation} \label{eqn_2ndConvergenceConstant_Upper}
		\expctn \squarebracketsbig{\norm{\bvltn{\epsilon}_{i,t}}^2}
		\leq
		\parenthesesbig{1-\kappa_{\max}^{-2}}^{t-N_R+1} \expctn \squarebracketsbig{\norm{\bvltn{\epsilon}_{i, N_R-1}}^2} 
		+ N_R \frac{ \kappa_{\max}^2}{F_{\min}^2}  \squarebracketsbig{1-\parenthesesbig{1-\kappa_{\max}^{-2}}^{t-N_R+1}} \sigma_n^2
		\end{equation}
		\begin{equation} \label{eqn_2ndConvergenceConstant_Lower}
		\expctn \squarebracketsbig{\norm{\bvltn{\epsilon}_{i,t}}^2}
		\geq
		\parenthesesbig{1-\xi_{\min}^{-2}}^{t-N_R+1} \expctn \squarebracketsbig{\norm{\bvltn{\epsilon}_{i, N_R-1}}^2} 
		+ N_R \frac{ \xi_{\min}^2}{F_{\max}^2}  \squarebracketsbig{1-\parenthesesbig{1-\xi_{\min}^{-2}}^{t-N_R+1}} \sigma_n^2
		\end{equation}	
	\end{lem}
	
	\begin{thm}[Partial Convergence] \label{thrm_PartialConstant} 
		WLOG, assume for any $i$, data generated by subsystem $i$ will all be assigned to candidate $i$. Let $\bvltn{\epsilon}_{i,t} = \bv{w}_i - \wht{\bv{w}}_{i,t}$ denote the estimation error of candidate $i$ at time $t$. Then $\forall i,t$ such that $r(i,t) \geq N_R$, we have
		\begin{equation} \label{eqn_untitle41}
		\expctn \squarebracketsbig{\norm{\bvltn{\epsilon}_{i,t}}^2}
		\leq
		\parenthesesbig{1 {-}\kappa_{\max}^{-2}}^{r(i,t)-N_R+1} \parenthesesbig{\norm{\bvltn{\epsilon}_{i,0}}^2 {+} \frac{(N_R-1) }{S_{\min}^2}} \\
		+ N_R \frac{ \kappa_{\max}^2}{F_{\min}^2}  \squarebracketsbig{1-\parenthesesbig{1-\kappa_{\max}^{-2}}^{r(i,t)-N_R+1}} \sigma_n^2
		\end{equation}
		\begin{equation} \label{eqn_untitle42}
		\expctn \squarebracketsbig{\norm{\bvltn{\epsilon}_{i,t}}^2}
		\geq
		\parenthesesbig{1-\xi_{\min}^{-2}}^{r(i,t)-N_R+1}  \frac{\sigma_n^2}{\phi_{\max}^2} \\
		+ N_R \frac{ \xi_{\min}^2}{F_{\max}^2}  \squarebracketsbig{1-\parenthesesbig{1-\xi_{\min}^{-2}}^{r(i,t)-N_R+1}} \sigma_n^2
		\end{equation}		
		If as $t \rightarrow \infty$, we have $r(i,t) \rightarrow \infty$, i.e. subsystem $i$ can dominate infinitely often, then as $t\rightarrow\infty$, we shall have
		\begin{equation} \label{eqn_untitle43}
		N_R \frac{\xi_{\min}^2}{F_{\max}^2}   \sigma_n^2
		\leq
		\expctn \squarebracketsbig{\norm{\bvltn{\epsilon}_{i,t}}^2}
		\leq
		N_R \frac{\kappa_{\max}^2}{F_{\min}^2}   \sigma_n^2
		\end{equation}	
	\end{thm}
	
	\subsection{Local Convergence Results} \label{subsctn_LocalConvrg}

	In this section, we present results regarding local convergence. Lemma \ref{lemma_noMisassignment} shows that when all candidates have accurate enough estimates, then the next assignment will be correct. Lemma \ref{lemma_ProbBoundforSingleSys}, which is derived from Lemma \ref{lemma_Supermartingale}, gives a lower bound on the probability that the estimates will stay accurate enough during the algorithm assuming assignments are correct. By Lemma \ref{lemma_ProbBoundforSingleSys} and Lemma \ref{lemma_noMisassignment}, we could obtain the local convergence result Theorem \ref{thrm_LocalConvergence}.
	
	For Lemma \ref{lemma_Supermartingale} to hold, we need a technical assumption given below to guarantee that estimation errors form a supermartingale, which allows us to use supermartingale maxima inequality to get the probability bound in Lemma \ref{lemma_Supermartingale}. Note that Assumption \ref{asmp_SNRUpperBound} is not mandatory for the algorithm to work, instead, it's solely for analysis purposes. Also note that if there is no noise, even though Assumption \ref{asmp_SNRUpperBound} fails, the estimation errors still form a supermartingale, which enables us to proceed with the analysis. Local convergence for the noiseless case is provided in Corollary \ref{crlry_LocalConvergenceWONoise}. 	
	
	\begin{assum}\label{asmp_SNRUpperBound}
		Assume there is an upper bound on the SNR: $\forall t, \frac{\norm{\bvltn{\phi}_t}}{ | n_t|} \leq S_{\max}$, which satisfies $S_{\max} \leq \kappa_{\max} S_{\min}$.
	\end{assum}

	\begin{lem} \label{lemma_Supermartingale}
		Assume Assumption \ref{asmp_SNRUpperBound} holds. With Setup(A), for $\forall t\geq N_R$ and some $\epsilon'>0$, we have
		\begin{equation} \label{eqn_supermartingaleMaximaIneq}
		P \parenthesesbig{\bigcap_{\tau = N_R}^t \curlybracketsbig{\norm{\bvltn{\epsilon}_{i, \tau}}^2 \leq \epsilon'^2} } \geq 1 - \frac{\expctn \squarebracketsbig{\norm{\bvltn{\epsilon}_{i, N_R-1}}^2}}{\epsilon'^2} 
		\end{equation}
	\end{lem}
	
	\begin{lem} \label{lemma_ProbBoundforSingleSys}
		Assume Assumption \ref{asmp_SNRUpperBound} holds. With Setup(A), for some $\epsilon'>0$, assume $\norm{\bvltn{\epsilon}_{i,0}} \leq \epsilon_0$ such that $\sqrt{N_R \parenthesesbig{\epsilon_0^2 + \frac{N_R}{S_{\min}^2}}} \leq \epsilon'$, then for $\forall t$ we have
		\begin{equation} \label{eqn_ProbBoundforSingleSys}
		P \parenthesesbig{\bigcap_{\tau = 1}^t \curlybracketsbig{\norm{\bvltn{\epsilon}_{i, \tau}}^2 \leq \epsilon'^2} } {\geq} 1 - 2 \sqrt{ \frac{N_R}{\epsilon'^2} \parenthesesbig{\epsilon_0^2 + \frac{N_R}{S_{\min}^2}} }
		\end{equation}
	\end{lem}
	
	\begin{lem} \label{lemma_noMisassignment}
		Let $\epsilon' = \frac{1}{2 \phi_{\max}} \parenthesesbig{\psi - \frac{n_{\max}}{\nu S_{\min}} - 3n_{\max}}$, $\alpha = 2$, and $\beta = 1$. Assume at time $t$, candidates are one-to-one $\epsilon'$-close to subsystems. WLOG, we could assume $\forall i, \norm{\bvltn{\epsilon}_{i, t-1}}\equiv\norm{\bv{w}_i - \wht{\bv{w}}_{i, t-1}} \leq \epsilon'$. Furthermore, we assume that all assignments prior to time $t$ are made correctly, i.e. $\forall s<t, \wht{\sysidx}_s = \sysidx_s$. Then at time $t$, we will also assign data correctly, i.e. $\wht{\sysidx}_t = \sysidx_t$.
	\end{lem}
	
	\begin{thm}[Local Convergence]  \label{thrm_LocalConvergence}
		Assume Assumption \ref{asmp_SNRUpperBound} holds. Let $\epsilon' = \frac{1}{2 \phi_{\max}} \parenthesesbig{\psi - \frac{n_{\max}}{\nu S_{\min}} - 3n_{\max}}$, $\alpha = 2$, and $\beta = 1$. Let $\bvltn{\epsilon}_{i,t} = \bv{w}_i - \wht{\bv{w}}_{i,t}$ denote the estimation error of candidate $i$ at time $t$. WLOG, assume $\forall i, \norm{\bvltn{\epsilon}_{i,0}} \leq \epsilon_0$ such that $\sqrt{N_R \parenthesesbig{\epsilon_0^2 + \frac{N_R}{S_{\min}^2}}} \leq \epsilon'$. Then $\forall i,t$ such that $r(i,t) \geq N_R$, with probability at least $1 - 2 \nsys \sqrt{ \frac{N_R}{\epsilon'^2} \parenthesesbig{\epsilon_0^2 + \frac{N_R}{S_{\min}^2}} }$, we have the following results: (i). We can correctly identify the switching sequence, i.e. $\forall t, \wht{\sysidx}_t = \sysidx_t$. In another way, $\forall i,t $, $\curlybrackets{\bvltn{\phi}_t, y_t}$ from subsystem $i$ will be assigned to candidate $i$. (ii). Results for \eqref{eqn_untitle41}, \eqref{eqn_untitle42} and \eqref{eqn_untitle43} will hold.
	\end{thm}
	
	\begin{cor}[Local Convergence Without Noise]  \label{crlry_LocalConvergenceWONoise} Let $n_t  = 0$, i.e. there is no noise. Let $\epsilon' {=} \frac{\psi}{2 \phi_{\max}}$, $\alpha {=} 2$, and $\beta {=} 1$. Let $\bvltn{\epsilon}_{i,t} {=} \bv{w}_i - \wht{\bv{w}}_{i,t}$ denote the estimation error of candidate $i$ at time $t$, and assume $\forall i, \norm{\bvltn{\epsilon}_{i,0}} \leq \epsilon_0$ such that $\sqrt{N_R \epsilon_0^2} \leq \epsilon'$.
		
	Then $\forall i,t$ such that $r(i,t) \geq N_R$, with probability at least $1 - 2 \nsys \sqrt{ \frac{N_R}{\epsilon'^2} \epsilon_0^2 }$, we have the following results: (i). We can correctly identify the switching sequence, i.e. $\forall t, \wht{\sysidx}_t = \sysidx_t$. In another way, $\forall t, \forall i$, $\curlybrackets{\bvltn{\phi}_t, y_t}$ from subsystem $i$ will be assigned to candidate $i$. (ii). we have the following convergence results: $\forall i,t$ such that $r(i,t) \geq N_R$
	\begin{equation} \label{eqn_untitle44}
	\expctn \squarebracketsbig{\norm{\bvltn{\epsilon}_{i,t}}^2}
	\leq
	\parenthesesbig{1-\kappa_{\max}^{-2}}^{r(i,t)-N_R+1} \norm{\bvltn{\epsilon}_{i,0}}^2
	\end{equation}
	If as $t {\rightarrow} \infty$, we have $r(i,t) {\rightarrow} \infty$, i.e. subsystem $i$ can dominate infinitely often, then as $t{\rightarrow}\infty$, we have $\expctn \squarebracketsbig{\norm{\bvltn{\epsilon}_{i,t}}^2} {=} 0$.
	\end{cor}

	
	\section{Discussions and Extensions} \label{sctn_Extensions}
	
	\subsection{Poles, Condition Number, and Convergence Rate}\label{subsctn_PolesvsConvry}
	
	Systems with poles close to the unit circle are not preferable as they are close to be unstable. In algorithm convergence analysis, Hessian matrix or objective function with large condition number is usually not preferable as the convergence rate tends to get small. In this section, we will show how these two facts meet consistently in our algorithm. That is, as the system poles getting closer to the unit circle, the condition number of Hessian matrix will get larger, and the convergence rate of upper bound in \eqref{eqn_untitle41} will get smaller. 

	To study the convergence rate, it suffices to study a single subsystem without any switching. We drop the subsystem subscript, and let $\bvltn{\epsilon}_t = \bv{w} - \wht{\bv{w}}_t$ denote the estimation error. Since the goal of this section is to provide insight into the relations between poles, condition number, and convergence rate, so several steps involve approximation. And when study how poles affect the condition number, we only consider a toy system with order 3, since it is challenging to find nice analytical expressions for systems with higher order. 

	First we consider how condition number influences convergence the rate of upper bound in \eqref{eqn_untitle41}.

	\subsubsection{Condition Number vs. Convergence Rate} \label{subsubsctn_CondvsConvrg}
	The expression for single ARX system is given by $y_t = \sum_{j=1}^{n_a} a_j y_{t-j} + \sum_{k=1}^{n_c} c_k u_{t-k} + n_t = \bv{w}^\T \bvltn{\phi}_t + n_t$ following our notations in Section \ref{sctn_ProblemFormulation}. In Assumption \ref{asmp_singularvalue}, we have $\singval_{\min}^2 I_n \preceq \sum\nolimits_{t\in S} \bvltn{\phi}_t \bvltn{\phi}_t^\T \preceq \singval_{\max}^2 I_n$. We will see this equation is related to the correlation matrix $\bv{R}\equiv \expctn[\bvltn{\phi}_t \bvltn{\phi}_t^\T]$, if it exists.

	In \cite{simon_s._haykin_[simon_2003}, we could know for the ARX system given above, if (i) poles of system are within the unit circle, and (ii) noise is white Gaussian and input is wide-sense stationary, then there exists $\bv{R}$ such that $\lim_{t\rightarrow \infty} \expctn[\bvltn{\phi}_t \bvltn{\phi}_t^\T] = \bv{R}$.

	When $N_R$ is large, according to law of large numbers, equation \eqref{eqn_untitle52} and the result above, we have $\splitatcommas{ \lim_{\min(S)\rightarrow \infty} \sum\nolimits_{t\in S} \bvltn{\phi}_t \bvltn{\phi}_t^\T \approx N_R \bv{R} }$. We let $\lambda_{\max}, \lambda_{\min}$ denote the maximum and minimum eigenvalue of $\bv{R}$. Now dropping the ``$\lim$'' and replace ``$\approx$'' with ``$=$'', we could get $N_R \lambda_{\min} I_n \preceq \sum\nolimits_{t\in S} \bvltn{\phi}_t \bvltn{\phi}_t^\T \preceq N_R \lambda_{\max} I_n$. So $N_R \lambda_{\min}$ and $N_R \lambda_{\max}$ are equivalent to $\singval_{\min}^2$ and $\singval_{\max}^2$ defined in Assumption \ref{asmp_singularvalue}. Then according to Lemma \ref{lemma_boundsonNorm}, we could have $\kappa_{\max} = \sqrt{(n{-}1) \lambda_{\max} / \lambda_{\min} {+} 1 }$ and $\xi_{\min} = \sqrt{(n{-}1) \lambda_{\min} / \lambda_{\max} {+} 1 }$. So, for the asymptotic convergence upper bounds in \eqref{eqn_2ndConvergenceConstant_Upper}, \eqref{eqn_untitle41}, and \eqref{eqn_untitle44} which all involve $\kappa_{\max}$, when the condition number of $\bv{R}$, $\lambda_{\max} / \lambda_{\min}$, increases, $\kappa_{\max}$ will increase, and the convergence rate in upper bounds will decrease.

	\subsubsection{Poles vs. Condition Number} \label{subsubsctn_PolevsCond}
	We consider a toy example of system with order 3: $y_t = a_1 y_{t-1} + a_2 y_{t-2} + c_1 u_{t-1} + n_t = \bv{w}^\T \bvltn{\phi}_t + n_t$ where $\bv{w} \equiv [a_1, a_2, c_1]$ and $\bvltn{\phi}_t \equiv [y_{t-1}, y_{t-2}, u_{t-1}]$. We assume all poles are within the unit circle, $u_t \sim \N(0, \sigma_u^2)$, $n_t \sim \N(0, \sigma_n^2) $, $u_t \perp n_t$, $u_t \perp u_s$, $n_t \perp n_s, \forall t, s$, and $\sigma_u \gg \sigma_n$. Following \cite{simon_s._haykin_[simon_2003}, we have
	\begin{equation}
	\bv{R} = \expctn \squarebracketsbig{\bvltn{\phi}_t \bvltn{\phi}_t^\T}=
	\begin{bmatrix}
	r(0) & r(1) & 0\\ 
	r(1) & r(0) & 0\\ 
	0 &  0& \sigma_u^2
	\end{bmatrix}
	\end{equation}
	where $r(0), r(1)$ can be computed by solving
	\begin{equation}
	\begin{bmatrix}
	r(0)\\ 
	r(1)\\ 
	r(2)
	\end{bmatrix}
	=
	\begin{bmatrix}
	1 & -a_1 & -a_2\\ 
	-a_1 & 1{-}a_2 & 0\\ 
	-a_2 & -a_1 & 1
	\end{bmatrix}^{-1}	
	\begin{bmatrix}
	\sigma_n^2 {+} c_1 \sigma_u^2\\ 
	0\\ 
	0
	\end{bmatrix}
	\end{equation}
	So, we have
	\begin{equation} \label{eqn_R}
	\bv{R} = \expctn \squarebracketsbig{\bvltn{\phi}_t \bvltn{\phi}_t^\T}=
	\begin{bmatrix}
	(a_2-1)c & (-a_1)c & 0\\ 
	(-a_1)c & (a_2-1)c & 0\\ 
	0 &  0& \sigma_u^2
	\end{bmatrix}
	\end{equation}
	where $c {=} \frac{\sigma_n^2 {+} c_1 \sigma_u^2}{(a_2{+}1)(a_1{+}a_2{-}1)(a_1{-}a_2{+}1)}$. We will drop $\sigma_n^2$ in the following computation as $\sigma_u \gg \sigma_n$. The eigenvalues of $\bv{R}$ are given by
	\begin{align} 
	\lambda_1 &= -\frac{c_1 \sigma_u^2}{(a_2+1)(a_1+a_2-1)} \\
	\lambda_2 &= \frac{c_1 \sigma_u^2}{(a_2+1)(a_1-a_2+1)} \\
	\lambda_3 &= \sigma_u^2
	\end{align}
	Note that the poles $p_1,p_2$ satisfy $p_1 + p_2 = a_1$ and $p_1 p_2 = -a_2$, so we have		
	\begin{align} 
	\lambda_1 &= \frac{c_1 \sigma_u^2}{(1-p_1 p_2)(1-p_1)(1-p_2)} \\
	\lambda_2 &= \frac{c_1 \sigma_u^2}{(1-p_1 p_2)(1+p_1)(1+p_2)} \\
	\lambda_3 &= \sigma_u^2
	\end{align}
	Let $c_1 \geq 1$, since $p_1, p_2 < 1$, we can see the condition number of $\bv{R}$ will have the following lower bound
	\begin{equation}
	\frac{\lambda_{\max}}{\lambda_{\min}} \geq \frac{\lambda_1}{\lambda_3} = \frac{1}{(1-p_1 p_2)(1-p_1)(1-p_2)}
	\end{equation}
	It's easy to see as poles get closer to the unit circle, this lower bound will get larger and the condition number is likely to increase as well.

	\subsubsection{Poles vs. Convergence Rate}
	Finally, by combining the two results we just showed, we could see as the system poles getting closer to unit circle, the convergence rate of upper bound in \eqref{eqn_untitle41} will decrease.
	
	There are two comments regarding this conclusion. (i) Even though this result only involves the rate of \emph{upper bound}, empirical results show the true convergent rate follow accordingly; (ii) Our algorithm favors stable system which is a little counterintuitive as unstable system tends to have higher SNR.

	\subsection{Unbounded Noise and Monte Carlo Method} \label{subsctn_unboundedNoise}
	
	Note that we compute the error upper bound $\epsilon_{\wht{\sysidx}_t, t}^u$ in Line \ref{algline_untitled13} of Algorithm \ref{alg_computeUpperbound} by finding the maximum $\norm{A \bv{n} - \bv{b}}$ from cube vertices $V$ defined by the noise magnitude upper bound $n_{\max}$. However, if $n_{\max}$ is unknown or the noise itself is unbounded, e.g. Gaussian, Algorithm \ref{alg_computeUpperbound} is not applicable to evaluate $\epsilon_{\wht{\sysidx}_t, t}^u$. In this case, if we could have samples of noise instead, an alternative approach is to use Monte Carlo method to evaluate $\epsilon_{\wht{\sysidx}_t, t}^u$. Specifically, if we have $N_t$ samples of noise vector $\bv{n}_t$ (defined in the proof for Theorem \ref{thrm_validUpperbound}) given by $\curlybrackets{\bv{n}_t^{(i)}}_{i=1}^{N_t}$, we could let $\epsilon_{\wht{\sysidx}_t, t}^u = \max \norm{A \bv{n}_t^{(i)} - \bv{b}}$. Due to the Monte Carlo nature, this is not necessarily a valid upper bound. In another way, the result in Theorem \ref{thrm_validUpperbound} doesn't hold, i.e. $\epsilon_{\wht{\sysidx}_t, t}^u < \norm{\bvltn{\epsilon}_{\wht{\sysidx}_t, t}}$. Practically, algorithm still has satisfactory performance when using this Monte Carlo method, but theoretically, this may not guarantee local convergence since local convergence result Theorem \ref{thrm_LocalConvergence} implicitly Theorem \ref{thrm_validUpperbound}, i.e. $\epsilon_{\wht{\sysidx}_t, t}^u \geq \norm{\bvltn{\epsilon}_{\wht{\sysidx}_t, t}}$ , to hold for every time step.

	If we prefer the theoretical guarantees to practical implementation, by subtly designing the number of Monte Carlo samples $N_t$ at time $t$, there could be some probability guarantee to ensure $\epsilon_{\wht{\sysidx}_t, t}^u$ is a valid upper bound at every time step.

	For ease of illustration, we assume there is only one subsystem, then we could drop the subsystem index subscript, and replace $\epsilon_{\wht{\sysidx}_t, t}^u$ with $\epsilon_{t}^u$, $\bvltn{\epsilon}_{\wht{\sysidx}_t, t}$ with $\bvltn{\epsilon}_{t}$. And we assume the Monte Carlo method starts at time $1$. Then we have the following theorem:

	\begin{thm} \label{thrm_IncreaseMonteCarlo}
		If we use Monte Carlo method above to compute $\epsilon_{t}^u$, for some $\zeta_1, \zeta_2 \in (0,1)$, let $N_t \geq \frac{\zeta_2 t}{2\zeta_1^{2t}}$,  then
		\begin{equation} \label{eqn_untitle48}
			P\parenthesesbig{P\parenthesesbig{\bigcap_{t=1}^\infty \curlybracketsbig{\norm{\bvltn{\epsilon}_{t}} {\leq} \epsilon_{t}^u}  } \geq 1{-}\frac{\zeta_1}{1-\zeta_1}} 
			\geq 1- \frac{\exp(-\zeta_2)}{1- \exp(-\zeta_2)}
		\end{equation}
		In another word, the probability that every $\epsilon_{t}^u$ is a valid upper bound is large with a large probability.
	\end{thm}

	The proof for this theorem is again in the appendices. We can immediately see from Theorem \ref{thrm_IncreaseMonteCarlo} that in order to make the probabilities large, the number of Monte Carlo samples need to increase exponentially with respect to time, which makes implementation intractable when time is long.
	
	\subsection{Extension to MIMO Case}

	So far we have been a considering SISO system in \eqref{eqn_ARXSystem}, where all the $y_t$ and $u_t$ are scalars. However, our algorithm can be applied to MIMO systems with some transformation of the system equation, and we will provide a potential direction in this section.

	Let $\bv{y}_t \in \R^{n_y}, \bv{u}_t \in \R^{n_u}$, then the MIMO SARX system is given by
	\begin{equation}
		\bv{y}_t = \sum_{j=1}^{n_a} \bv{A}_j(\sysidx_t) \bv{y}_{t-j} + \sum_{k=1}^{n_c} \bv{C}_k(\sysidx_t) \bv{u}_{t-k} + \bv{n}_t
	\end{equation}
	where $\curlybrackets{\bv{A}_j(\sysidx_t)}_{j=1}^{n_a}$, $\curlybrackets{\bv{C}_j(\sysidx_t)}_{j=1}^{n_c}$ are the parameters of subsystem $\sysidx_t$. Let $\splitatcommas{ \bv{W}_{\sysidx_t} = [\bv{A}_1(\sysidx_t), \dots, \bv{A}_{n_a}(\sysidx_t), \bv{C}_1(\sysidx_t), \dots \bv{C}_{n_c}(\sysidx_t) ]^\T }$, $\splitatcommas{ \bvltn{\phi}_t = [\bv{y}_{t-1}^\T, \dots,\bv{y}_{t-n_a}^\T, \bv{u}_{t-1}^\T, \dots, \bv{u}_{t-n_c}^\T]^\T }$. Let $\bv{w}_{\sysidx_t, i}$ denote the $i$th column of $\bv{W}_{\sysidx_t}$, and let $y_{t,i}, n_{t,i}$ denote the $i$th element in $\bv{y}_t$ and $\bv{n}_t$. Then the MIMO system can be broken into a set of equations: $\forall i \in [n_y]$,
	\begin{equation}
		y_{t,i} = \bv{w}_{\sysidx_t, i}^\T \bvltn{\phi}_t + n_{t,i}
	\end{equation}
		which has the same form as \eqref{eqn_ARXSystem}. So we could modify our algorithm to estimate each $\bv{w}_{\sysidx_t, i}$ in a parallel way, then combine them to estimate $\bv{W}_{\sysidx_t}$.

	\subsection{Multiple $N_C$'s and Forgetting Factor}
	
	\subsubsection{Multiple $N_C$'s}
	Note that in Algorithm \ref{alg_computeUpperbound}, we have window variables $\bvltn{\Phi}^C_{i,t} \in \R^{n \x N_C}, \wht{\bv{W}}^C_{i,t} \in \R^{n \x N_C}, \bv{h}^C_{i,t} \in \R^{N_C}$ for some window length $N_C$ to compute the error upper bound $\epsilon_{\wht{\sysidx}_t, t}^u$. Theorem \ref{thrm_validUpperbound} says  when all data stored in the window are from the same subsystem, then $\epsilon_{\wht{\sysidx}_t, t}^u$ will be a valid error upper bound with respect this subsystem. However, if there enters some \emph{outlier} data (data generated by subsystem that is different from the subsystem that generates the majority of data in the window variables), $\epsilon_{\wht{\sysidx}_t, t}^u$ computed using window variables might be an invalid upper bound. If the window length is too large, since the window is sliding, the effect of outlier will stay a longer time, but the correction effect of the majority of the inlier data might reduce the effect of outlier. On the contrary, if the window length is too small, the effect of outlier will quickly vanish, but the correction effect from the inlier data will reduce as well and we may have even worse $\epsilon_{\wht{\sysidx}_t, t}^u$ during the stay of outlier. 

	Practically, we could use multiple $N_C$'s and corresponding window variables. Each set of window variables compute $\epsilon_{\wht{\sysidx}_t, t}^u$ separately, and we pick the maximum of them as the final decision. In this way, the disadvantages of large and small window lengths might cancel out each other thus making $\epsilon_{\wht{\sysidx}_t, t}^u$ more robust to misassignment.

	\subsubsection{Forgetting Factor}
	One interesting fact about our algorithm is, in Line \ref{algline_PickColumn} of Algorithm \ref{alg_main}, instead of using the latest data, we pick randomly from previous data to update the estimates. This idea is initially proposed in \cite{strohmer2009randomized}. The reason we incorporate this randomization into the algorithm is to acquire the asymptotic convergence result via Assumption \ref{asmp_singularvalue}, Lemma \ref{lemma_boundsonNorm}, Lemma \ref{lemma_NoiseData}, and Lemma \ref{lemma_boundonExpectation}.
	
	If no randomization scheme is utilized, the algorithm on a single subsystem is equivalent to the Kaczmarz algorithm or the normalized least mean squares (NLMS) algorithm in \cite{simon_s._haykin_[simon_2003}. This type of algorithm, however, does not have satisfactory convergence results yet. One linear convergence result provided in \cite{simon_s._haykin_[simon_2003} is valid only when the step size in estimate update is very small, which makes it little practical use. The difficulty to derive nice convergence results is that nearby data $\bvltn{\phi}_{t}$ could be highly correlated, which can be seen from the definition, and updating the estimate with data in chronological order aggravates the situation. The randomized scheme picks data for update randomly and independently, which brings independence into the algorithm and makes analysis tractable. 	
	
	Empirically, if we don't incorporate the random selection in Line \ref{algline_PickColumn}, and always use latest data as Line \ref{algline_untitled4}, the performance can sometimes be slightly better. This is potentially because when sampling previous data vectors, it's likely that we sample one data multiple times possibly due to its large norm, and, generally speaking, previously used data may not provide as much information as some new data.

	One potential way to balance between establishing theoretical results and exploiting new data is to incorporate a forgetting factor $\gamma$. Specifically, in Line \ref{algline_PickColumn}, we sample data according to the following distribution
	\begin{equation}
		P(l_t=i)
		\begin{cases}
		& \gamma \norm{\bvltn{\Phi}^R_{{\wht{\sysidx}_t}, t}[: \, , \, i]}^2 \frac{1}{F} \text{ if } i=N_R \\ 
		& (1-\gamma) \norm{\bvltn{\Phi}^R_{{\wht{\sysidx}_t}, t}[: \, , \, i]}^2 \frac{1}{F} \text{ if }\text{ if } i<N_R 
		\end{cases}
	\end{equation}
	where $\gamma>0.5$, and $F = \gamma \norm{\bvltn{\Phi}^R_{{\wht{\sysidx}_t}, t}[: \, , \, N_R]}^2 + (1-\gamma) \sum_{i=1}^{N_R-1} \norm{\bvltn{\Phi}^R_{{\wht{\sysidx}_t}, t}[: \, , \, i]}^2$ is the normalization factor. With this distribution, we can see the probability of choosing the latest data ($l_t = N_R$) is larger compared with the distribution in Algorithm \ref{alg_main}. As $\gamma$ gets closer to $1$, we are more likely to sample the latest data.

	As for the convergence result, it suffices to only consider how the building block lemmas will change with this new distribution, and the main theorems will follow these lemmas. In the building block lemmas, only the expectations in Lemma \ref{lemma_NoiseData} (\ref{lemmaresult_3}), and Lemma \ref{lemma_boundonExpectation} involve the data sampling process. With the new sampling distribution, it's not difficult to see \eqref{eqn_untitle101} and \eqref{eqn_untitle102} will become

	\begin{equation}
	\tilde{\gamma}^{-1}\frac{N_R}{F_{\max}^2} \sigma_n^2
	\leq
	\expctn \squarebracketsbig{\frac{n_{r_t(l_t)}^2}{\norm{\bvltn{\phi}_{r_t(l_t)}}^2}} 
	\leq  
	\tilde{\gamma} \frac{N_R}{F_{\min}^2} \sigma_n^2
	\end{equation}
		
	\begin{equation}
	\tilde{\gamma}^{-1} \kappa_{\max}^{-2} \expctn [\norm{\bv{z}}^2 ]
	{\leq }
	\expctn \squarebracketsbig{\parenthesesbig{\frac{\bvltn{\phi}_{r_t(l_t)}^\T \bv{z} }{\norm{\bvltn{\phi}_{r_t(l_t)}}}  }^2} 
	{\leq} 
	\tilde{\gamma} \xi_{\min}^{-2} \expctn [ \norm{\bv{z}}^2 ]
	\end{equation}
	where $\tilde{\gamma} = \frac{\gamma}{1-\gamma}>1$. The rest of the lemmas, theorems, corollaries follow from these new results.

	
	\section{Numerical Results} \label{sctn_Experiments}
	
	In this section, we use simulation examples to evaluate the theoretical results as well as the performance of our algorithm.
	
	\subsection{Evaluation of Asymptotic Convergence Bounds}
	
	Since it is not convenient to visualize the convergence bounds for SARX system with multiple subsystems, and \eqref{eqn_2ndConvergenceConstant_Upper} and \eqref{eqn_2ndConvergenceConstant_Lower} give tighter performance than \eqref{eqn_untitle42} and \eqref{eqn_untitle43}, we will evaluate the lower and upper asymptotic convergence bounds in \eqref{eqn_2ndConvergenceConstant_Upper} and \eqref{eqn_2ndConvergenceConstant_Lower} on single ARX system by comparing the bounds with the actual convergence behavior.

	Consider a specific system
	\begin{equation}
	y_t = 0.7 y_{t-1} - 0.12 y_{t-2} + u_{t-1} + n_t
	\end{equation}
	where $n_t \sim \N(0, \sigma_n^2)$, $\sigma_n=10^{-4}$, $u_t \sim \N(0, 1)$. According to Section \ref{subsctn_PolesvsConvry}, the correlation matrix is given by  
	\begin{equation}
	R=
	\begin{bmatrix}
	1.67 & 1.04 & 0\\ 
	1.04 & 1.67 & 0\\ 
	0 & 0 & 1
	\end{bmatrix}
	\end{equation}
	, and its minimum and maximum eigenvalues are $\lambda_{\min}=0.63$ and $\lambda_{\max}=2.71$.

	Since it's difficult to have exact knowledge of $\kappa_{\max}$ and $\xi_{\min}$, we will use the approximate values defined in Section \ref{subsctn_PolesvsConvry}, i.e. $\kappa_{\max} = \sqrt{(n{-}1) \lambda_{\max} / \lambda_{\min} {+} 1 }$ and $\xi_{\min} = \sqrt{(n{-}1) \lambda_{\min} / \lambda_{\max} {+} 1 }$. And similarly, we could have $F_{\max} = \sqrt{n N_R \lambda_{\max}}$ and $F_{\min}= \sqrt{n N_R \lambda_{\min}}$. 

	We set $N_R = 10$ and simulation time horizon $T=1000$. To evaluate the expectation $\expctn \squarebracketsbig{\norm{\bvltn{\epsilon}_{i, t}}^2}$ in \eqref{eqn_2ndConvergenceConstant_Upper} and \eqref{eqn_2ndConvergenceConstant_Lower}, we run the algorithm $50$ times with different realizations of input $u_t$, noise $n_t$, and random data selection in Line \ref{algline_PickColumn} of Algorithm \ref{alg_main}, and take the average of estimation errors as the expectation. 

	\begin{figure}[h!]
		\begin{center}
			\includegraphics[width=7.5cm]{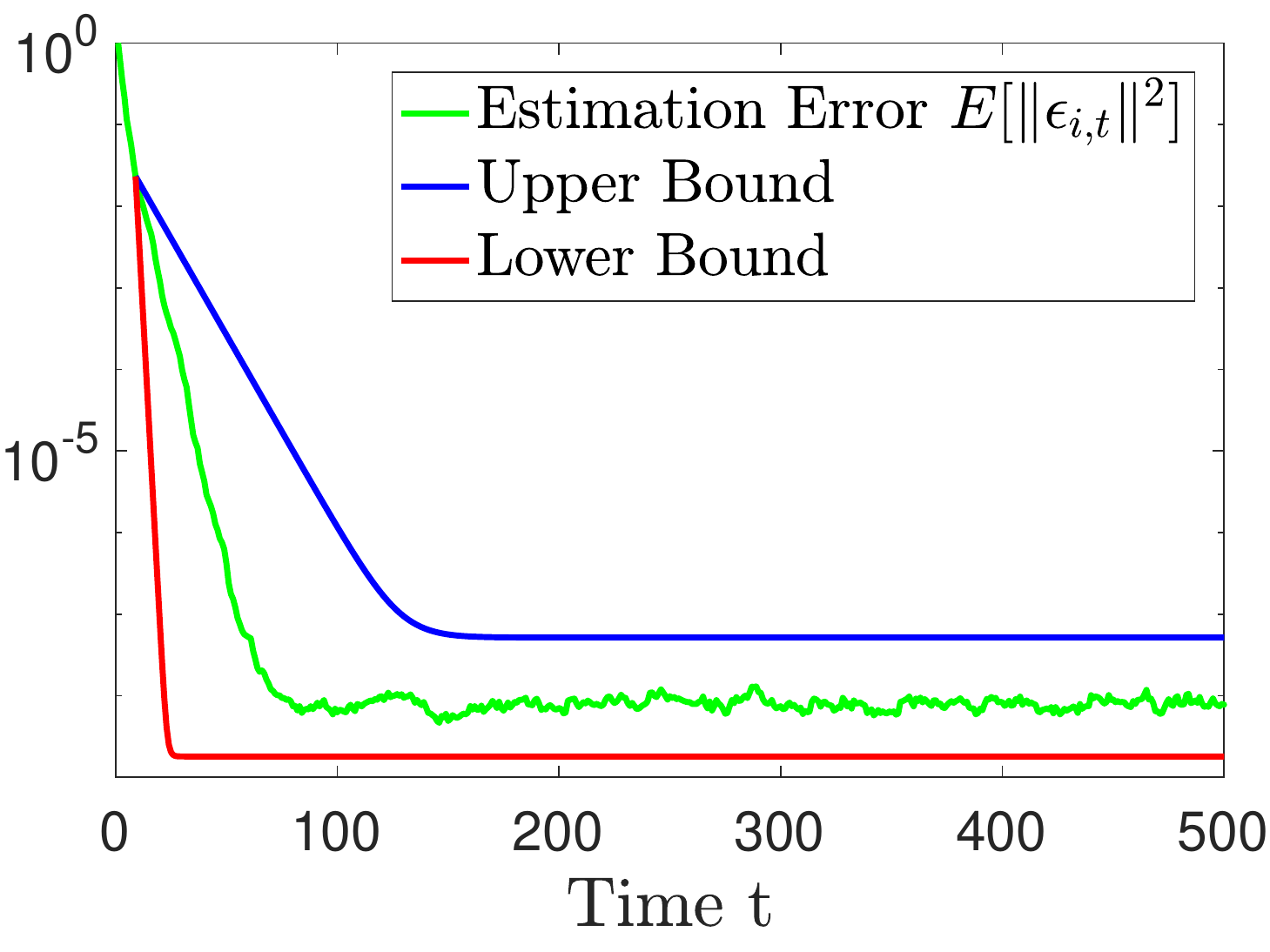}    
			\caption{\small Evaluation of convergence bounds} 
			\label{fig_BoundsTest}
		\end{center}
	\end{figure}

	The simulation results are given in Fig. \ref{fig_BoundsTest}. We can see the estimation error $\expctn \squarebracketsbig{\norm{\bvltn{\epsilon}_{i, t}}^2}$ can be successfully bounded by the upper and lower bound in \eqref{eqn_2ndConvergenceConstant_Upper} and \eqref{eqn_2ndConvergenceConstant_Lower}.

	\subsection{Robust Behavior of our Algorithm}
	
	\subsubsection{Single Realization Experiment}        
	First we evaluate our algorithm and compare it with the OBE algorithm in \cite{goudjil2016convergence} using SARX system given below
	\begin{itemize}
		\item Subsystem 1: $y_t = 0.2 y_{t-1} + 0.24 y_{t-2} + 2 u_{t-1} + n_t$
		\item Subsystem 2: $y_t = 0.7 y_{t-1} - 0.12 y_{t-2} + 1 u_{t-1} + n_t$
		\item Subsystem 3: $y_t = -1.4 y_{t-1} -0.53 y_{t-2} + 1 u_{t-1} + n_t$
		\item Subsystem 4: $y_t = 1.7 y_{t-1} -0.72 y_{t-2} + 0.5 u_{t-1} + n_t$
	\end{itemize}
	where $u_t {\sim} \N(0,1)$. $n_t$ follows $\N(0,\sigma_n^2)$ truncated to region $[-3\sigma_n, 3\sigma_n ]$ where $\sigma_n {=} 10^{-4}$, so noise is bounded with $n_{\max} {=} 3\sigma_n$.
	
	To fully evaluate the performance, we consider 3 different switching patterns of subsystems: (i) \emph{Slow Switching} (SS): subsystem 1 dominates from 1 to 500, subsystem 2 dominates from 501 to 1000, subsystem 3 dominates from 1001 to 1500, and subsystem 4 dominates from 1501 to 2000. (ii) \emph{Minimum Dwell Time} (MD): each subsystem dominates 30 time steps, and then the time it takes to switch to a new subsystem is a random variable following the geometric distribution with parameter $1/16$. When the subsystem switches, all subsystems are equally likely to be switched to, and after the switching, this process restarts again. (iii) \emph{Fast Switching} (FS): at every time step, every subsystem dominates with equal probabilities.
	
	In our algorithm, we set $N_R {=} 3, N_C{=}20, \alpha{=}4, \beta{=}3, \nu {=} 10^{-4}$,  and simulation time horizon $T=2000$. The candidates are initialized with standard multivariate Gaussian distribution. After the algorithms completes all $T$ time steps, we first relabel the candidates with a bijective mapping $h(\cdot): [\nsys]\rightarrow [\nsys]$ such that $\sum_{i \in [\nsys]} \norm{\bv{w}_i - \wht{\bv{w}}_{h(i),T}}$ is minimized. 
    \emph{In the following, the candidates are referring to the relabeled candidates.}
	
	We compute all the estimation errors, i.e. $\bvltn{\epsilon}_{i,t} = \bv{w}_i - \wht{\bv{w}}_{i, t}, \forall i, t$, which measure the distance between candidate $i$ and subsystem $i$ during the algorithm.
	
\begin{figure}[h!]
		\centering
		\captionsetup[subfloat]{captionskip=0pt, farskip=0pt}
		\subfloat[]{\includegraphics[width=2in]{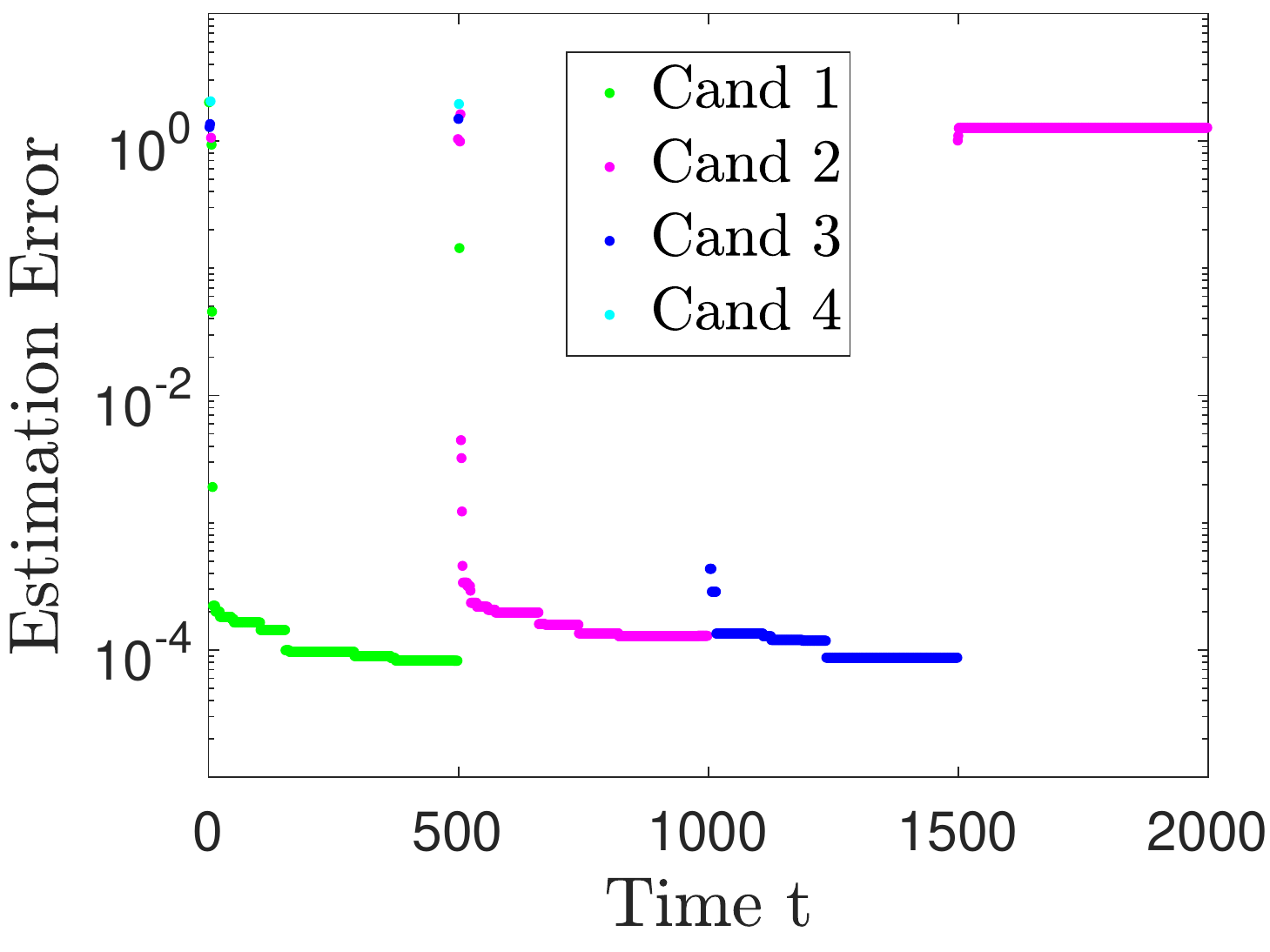} \label{1}} \ 
		\subfloat[]{\includegraphics[width=2in]{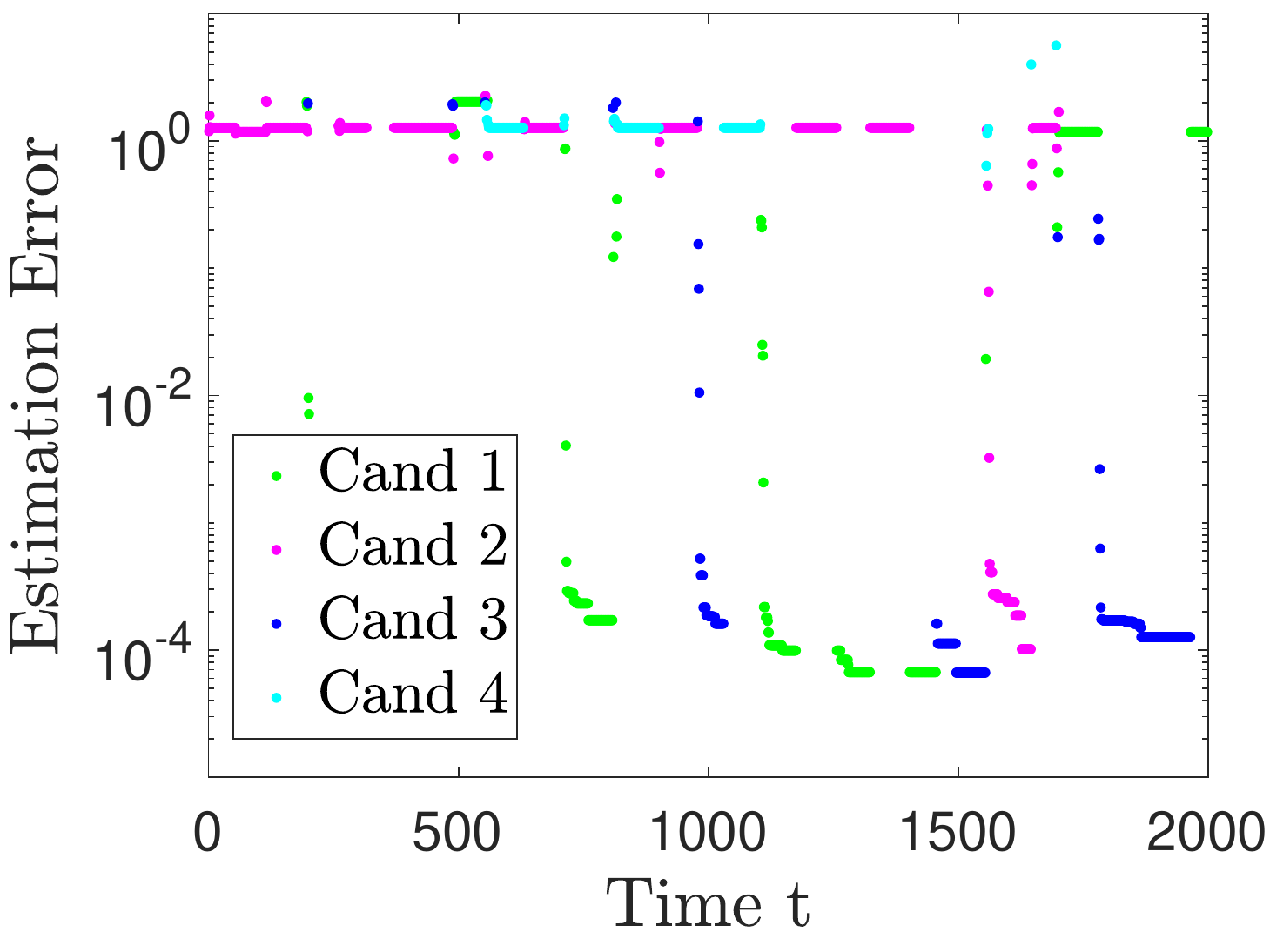} \label{3}} \ 
		\subfloat[]{\includegraphics[width=2in]{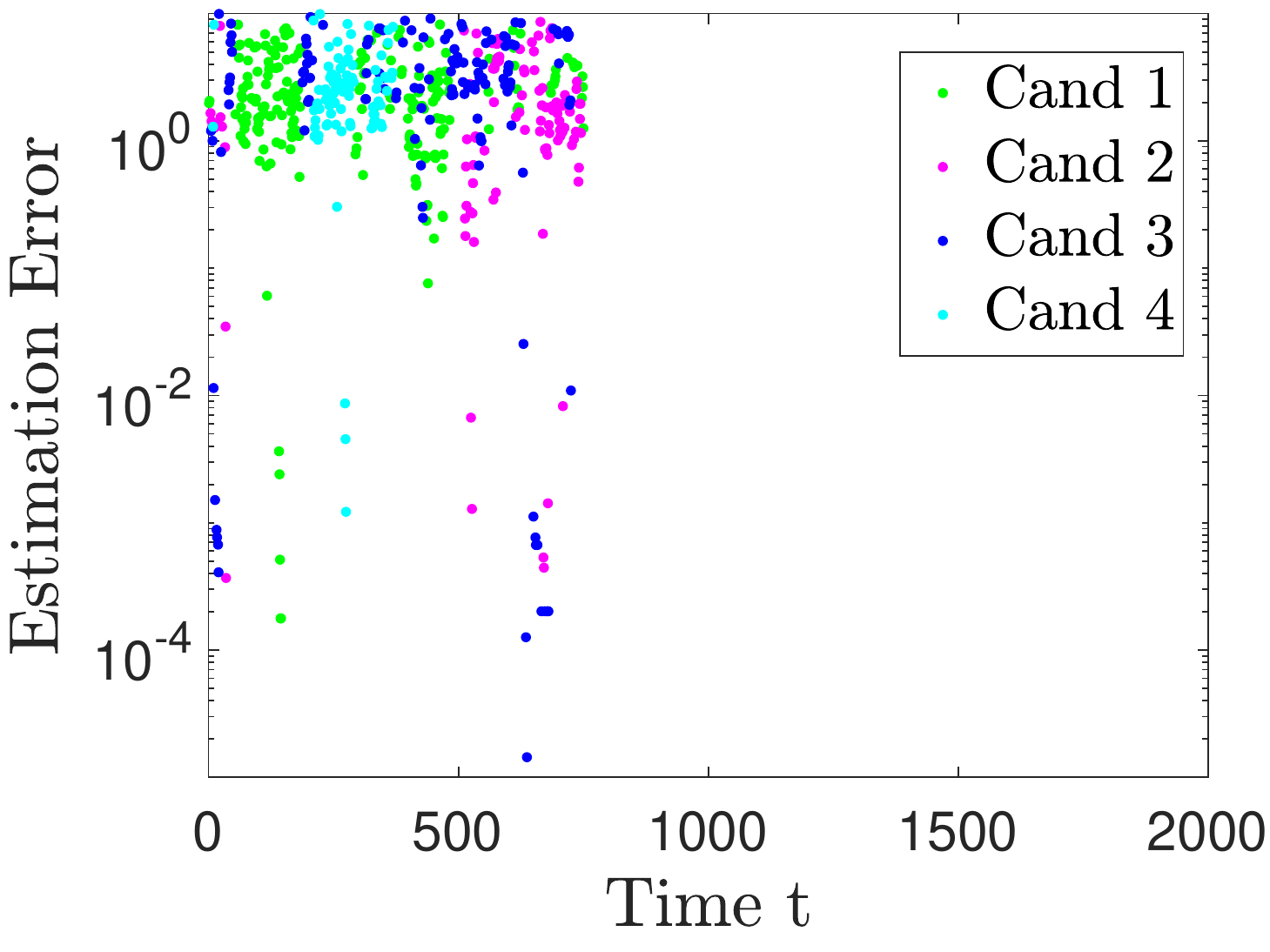} \label{5}} \\		
		\subfloat[]{\includegraphics[width=2in]{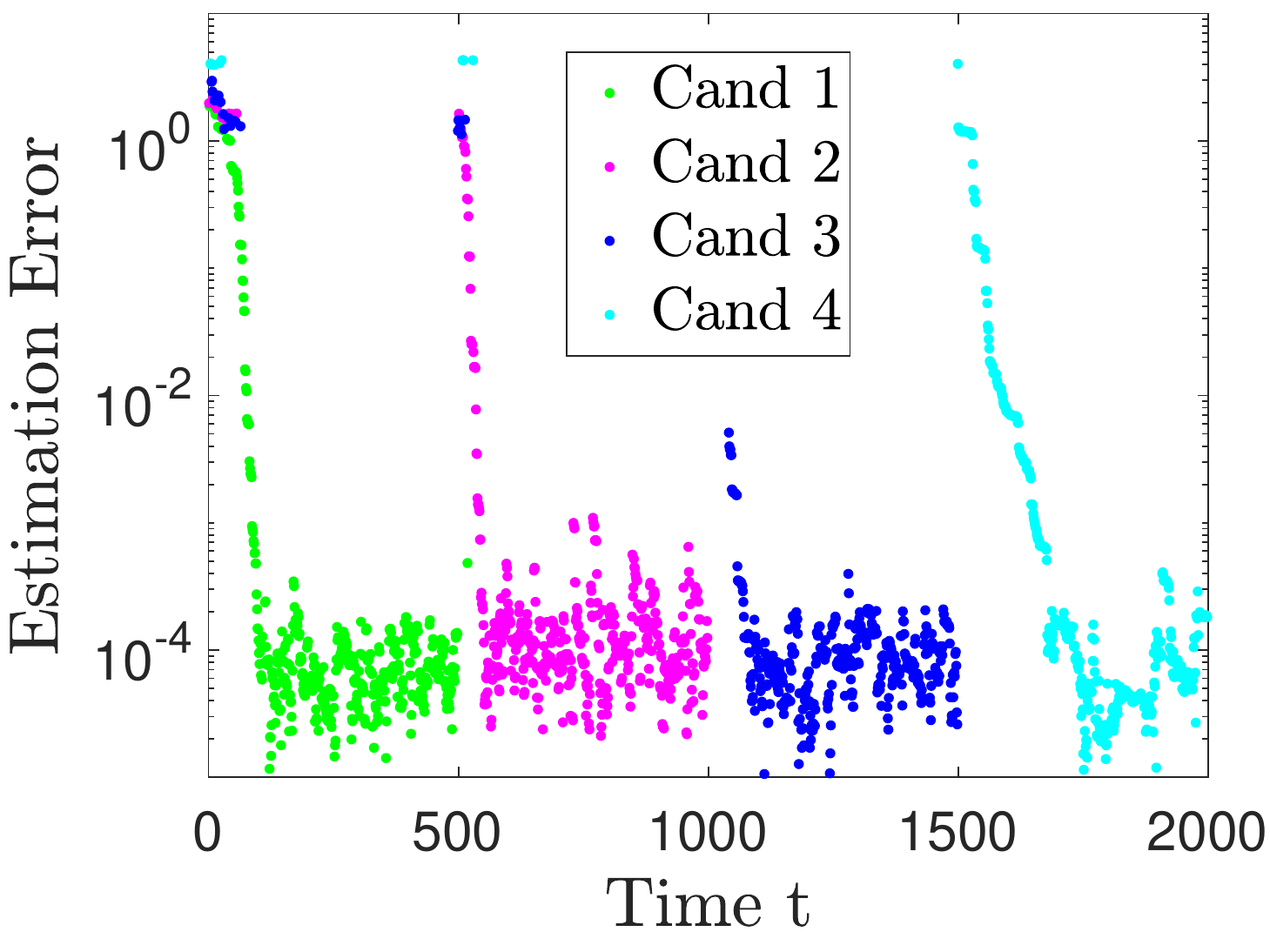} \label{2}} \
		\subfloat[]{\includegraphics[width=2in]{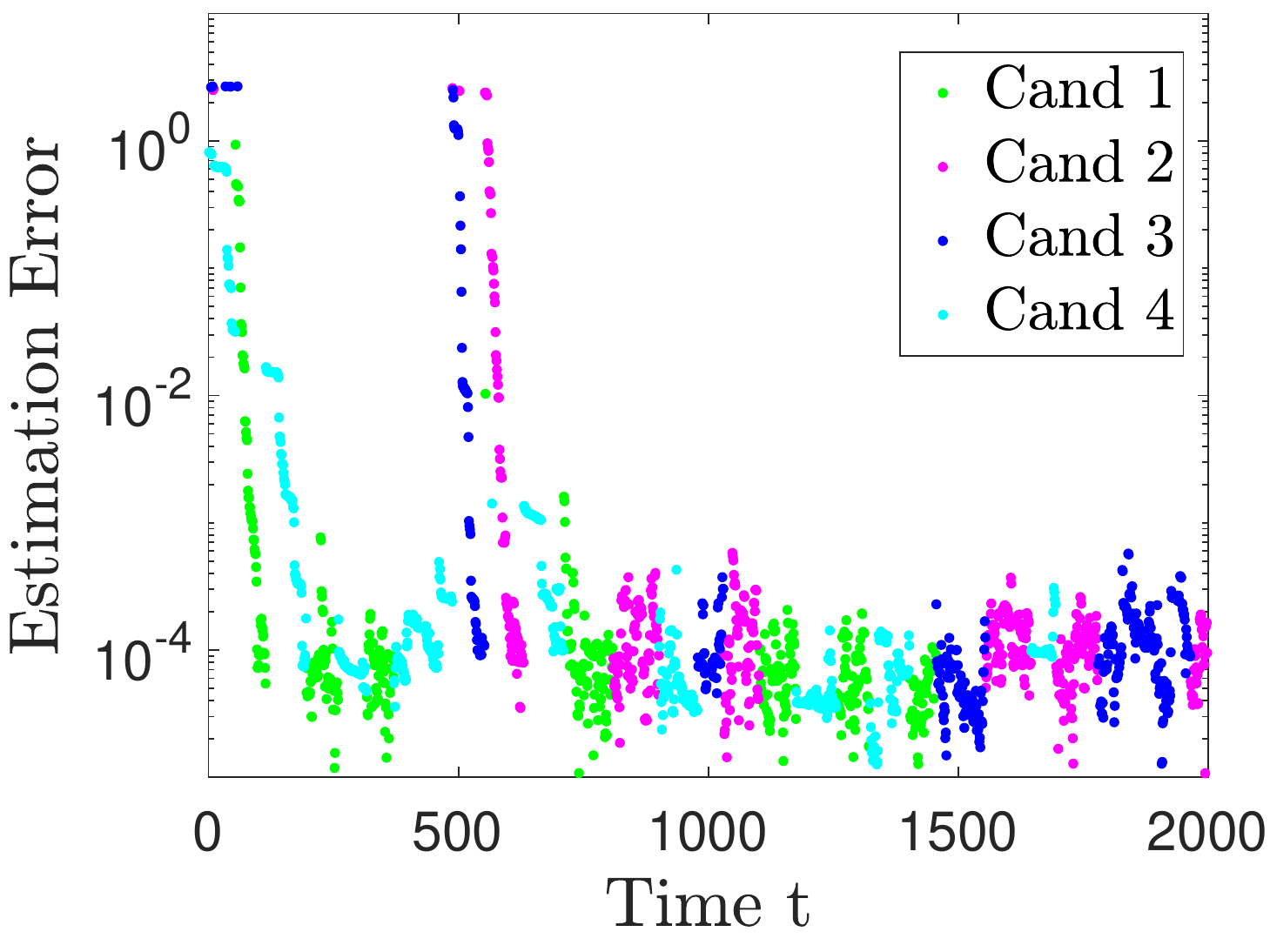} \label{4}} \
		\subfloat[]{\includegraphics[width=2in]{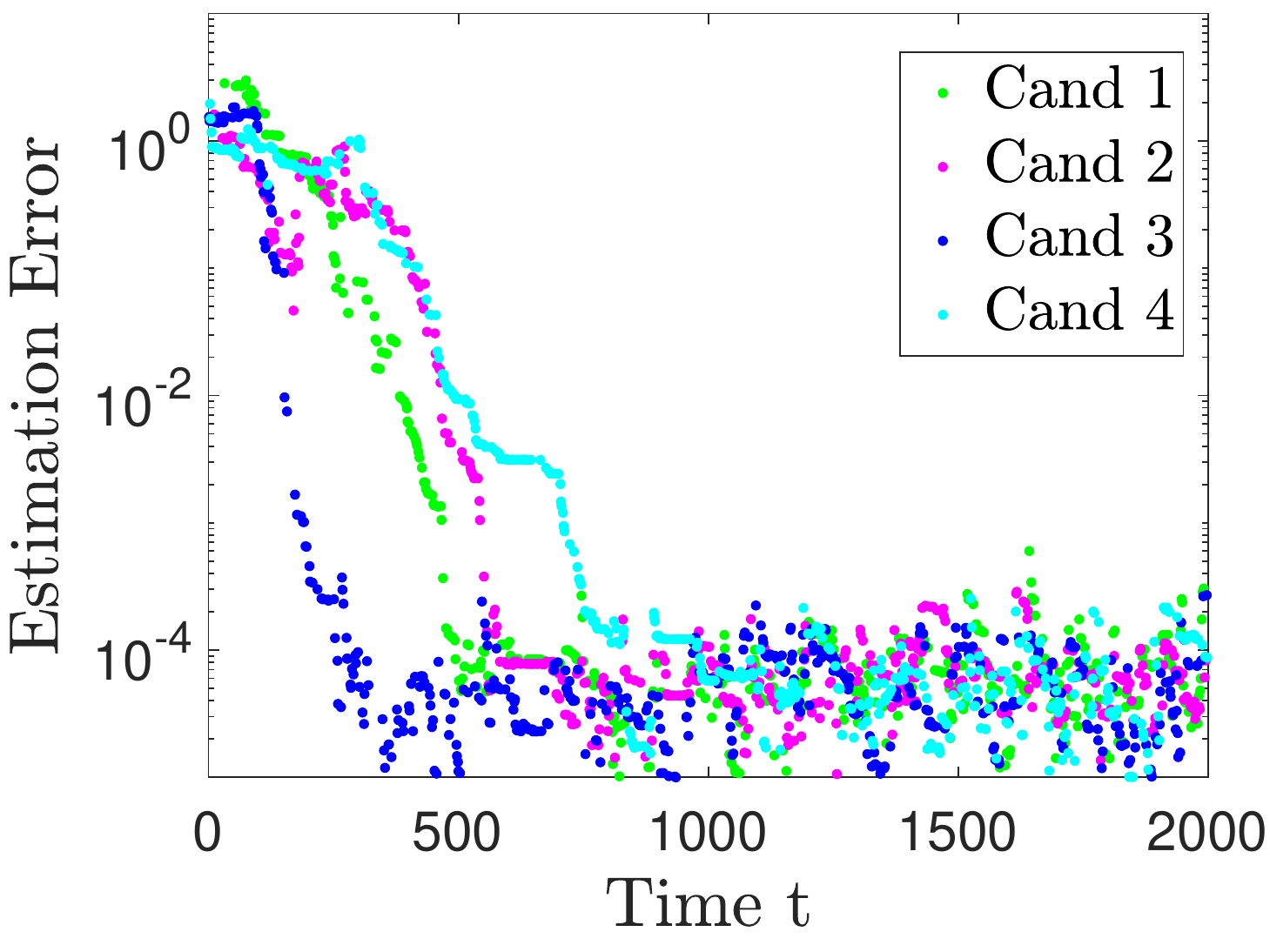} \label{6}} 
		\caption{\small Estimation errors of OBE algorithm and our algorithm: (a) SS-OBE; (b) MD-OBE; (c) FS-OBE; (d) SS-Ours; (e) MD-OBE; (f) FS-Ours}
		\label{fig_ConvrgComp}
	\end{figure}

	Fig. \ref{fig_ConvrgComp} depicts the simulation results. The dots in the plots represent each $\norm{\bvltn{\epsilon}_{\wht{\sysidx}_t,t}}, \forall t$, which means there is only one dot plotted for one time step. Different colors correspond to different candidates and corresponding true subsystems. For example, if there exists a blue dot at time $t=1400$, this means we assign data generated at time $t=1400$ to candidate 3, and current error between candidate 3 and subsystem 3 is given by the y-axis value of the dot.

	From the plots, we see that our algorithm converges more quickly than the OBE algorithm. Since none of the colors have a sharp increase in error, we could claim the phenomenon described in Section \ref{sctn_Drawback} is effectively avoided in these realizations. For the OBE algorithm, the performance is obviously worse: in the FS case, none of the candidates even converge, and the algorithm even stops halfway due to numerical instability. In the plots for SS-OBE, we can see the undesired phenomenon described in Section \ref{sctn_Drawback}: candidate 2 has converged to the vicinity of subsystem 2 from $t=501$ to $t=1000$, but after time $t\geq 1500$, the error goes large again. This is because we are assigning data generated by subsystem 4 to candidate 2, making candidate 2 move towards subsystem 4. From these plots, we see that our algorithm outperforms OBE algorithm for all switching patterns.

	\subsubsection{Multiple Realizations Experiment} 
	Since a single realization cannot comprehensively evaluates the performance, we further compare our algorithm with the OBE algorithm using multiple realizations. Specifically, we consider 9 experiment setups, given by all the combinations of switching patterns \{SS, MD, FS\} and noise level $\sigma_n \in \curlybrackets{10^{-1}, 10^{-2},  10^{-3}}$, and for each of the experiment setup, we run $M{=}100$ realizations. Each subsystem parameters are generated randomly in each realization: we first sample 2 real poles on $[-1, 1]$ uniformly, and then compute the parameters from the sampled poles. The rest of the setups, e.g. number and orders of subsystems, algorithm parameters, etc., follow the previous single realization experiment.
	
	For realization $i$, we define the two metrics: $\text{FE}(i) = \frac{1}{\nsys} \sum_{j=1}^{\nsys} \norm{\bvltn{\epsilon}_{j, T}}$ and $\text{CER}(i) = \frac{1}{T} \sum_{t=1}^T \indicator \{\sysidx_t \neq \wht{\sysidx}_{t} \}$. FE measures the final estimation error and CER is the classification error rate. Table \ref{tbl_table} lists the average FE and CER values over all 100 realizations. We could see, our algorithm exhibits better performance in each setup.
	\begin{table}[]
		\centering	
		\caption{\small Results for Multiple Realizations Experiment} \label{tbl_table}
		\renewcommand{\arraystretch}{1.3}
		\begin{tabular}{|l|l|l|l|l|}
			\hline
			& Ours & OBE & Ours & OBE \\ \hline
			& FE & FE & CER & CER\\ \hline
			SS, $10^{-1}$ 	& $8.4{\x}10^{-1}$ & $8.7{\x}10^{-1}$ 	& $56.3\%$ 	& $59.1\%$ \\ \hline
			SS, $10^{-2}$	& $2.8{\x}10^{-2}$ & $8.2{\x}10^{-1}$ 	& $22.1\%$ 	& $55.5\%$ \\ \hline
			SS, $10^{-3}$	& $9.0{\x}10^{-2}$ & $8.2{\x}10^{-1}$ 	& $8.35\%$  & $56.4\%$ \\ \hline
			MD, $10^{-1}$	& $4.3{\x}10^{-1}$ & $5.2{\x}10^{-1}$ 	& $47.5\%$ 	& $50.3\%$ \\ \hline
			MD, $10^{-2}$	& $4.0{\x}10^{-2}$ & $2.8{\x}10^{-1}$ 	& $11.3\%$	& $31.3\%$ \\ \hline
			MD, $10^{-3}$	& $9.4{\x}10^{-3}$ & $2.4{\x}10^{-1}$ 	& $4.91\%$ 	& $28.8\%$ \\ \hline
			FS, $10^{-1}$	& $2.6{\x}10^{-1}$ & $6.8{\x}10^{-1}$ 	& $39.3\%$ 	& $53.9\%$ \\ \hline
			FS, $10^{-2}$	& $6.0{\x}10^{-2}$ & $1.5{\x}10^{-1}$ 	& $11.7\%$ 	& $22.1\%$ \\ \hline
			FS, $10^{-3}$	& $5.8{\x}10^{-2}$ & $1.8{\x}10^{-1}$ 	& $8.93\%$ 	& $18.9\%$ \\ \hline
		\end{tabular}
		\label{my-label}
	\end{table}

	
	\section{Conclusions}
	In this paper, we introduced a robust algorithm to solve online switched system identification problem. Our algorithm follows the conventional two-step framework, but the modified assignment criterion leads to a more robust assignment process. After we assign the data to some candidate, we update the candidate estimate based on the idea of randomized Kaczmarz algorithm. We showed partial and local convergence results. The partial convergence result is: assuming there is no misassignment, then the estimation error converges geometrically to some quantity related to noise variance in the expectation square sense. The local convergence result is: assuming all candidates have good enough initialization, with some probability, it can be guaranteed that no misassignment will be made, and the estimation error will converge geometrically as in the partial result. Numerical results verify the asymptotic convergence bounds we developed, and shows the efficiency of our proposed algorithm in comparison with the existing OBE algorithm.
    
    For future work, there are several aspects that we would focus on. 
    \begin{itemize}
    	\item As for theories, we would seek to relax Assumption \ref{asmp_SNRUpperBound}, and analyze the local convergence in a more general setting. Also, we could relax even further to analyze the global convergence without good initialization requirement.
    	\item We plan to apply our algorithm to advanced and real world examples to further evaluate its applicability.
    	\item Our current algorithm finds the upper bound of estimation error by searching all the cube vertices $V$ defined in Algorithm \ref{alg_computeUpperbound}, which leads to heavy computation burden when $N_C$ is large. In the future, we would seek a way to estimate the error more efficiently without sacrificing theoretical guarantees.
    	\item Since so far we don't consider the case in which we have control over the system input, another interesting extension would be designing certain input, possibly closed-loop or open loop but with certain distribution, given which the system parameters can be learned faster.
    \end{itemize}

    
	\section*{Acknowledgement}
		The authors thank Yan Shuo Tan for suggesting the use of super-martingale theory, which proved to be crucial in the local convergence analysis.
	
	\bibliographystyle{abbrv}
	\bibliography{ifacconf}            

\begin{thebibliography}{10}

\bibitem{bako_identification_2011}
L.~Bako.
\newblock Identification of switched linear systems via sparse optimization.
\newblock {\em Automatica}, 47(4):668--677, Apr. 2011.

\bibitem{bako_recursive_2011}
L.~Bako, K.~Boukharouba, E.~Duviella, and S.~Lecoeuche.
\newblock A recursive identification algorithm for switched linear/affine
  models.
\newblock {\em Nonlinear Analysis: Hybrid Systems}, 5(2):242--253, May 2011.

\bibitem{bezruck2010application}
V.~Bezruck, Y.~N. Belov, O.~Voitovych, K.~Netrebenko, V.~Tikhonov, G.~Rudnev,
  G.~Khlopov, and S.~Khomenko.
\newblock Application of autoregressive model for recognition of meteorological
  objects.
\newblock In {\em Radar Symposium (IRS), 2010 11th International}, pages 1--3.
  IEEE, 2010.

\bibitem{cochrane2005time}
J.~H. Cochrane.
\newblock Time series for macroeconomics and finance.
\newblock {\em Manuscript, University of Chicago}, 2005.

\bibitem{goudjil2016convergence}
A.~Goudjil, M.~Pouliquen, E.~Pigeon, and O.~Gehan.
\newblock Convergence analysis of a real-time identification algorithm for
  switched linear systems with bounded noise.
\newblock In {\em Decision and Control (CDC), 2016 IEEE 55th Conference on},
  pages 2957--2962, 2016.

\bibitem{simon_s._haykin_[simon_2003}
S.~Haykin and B.~Widrow, editors.
\newblock {\em [{{Simon Haykin}}] {{Least}}-Mean-Square Adaptive Filters}.
\newblock Wiley series in adaptive and learning systems for signal processing,
  communication, and control. {Wiley-Interscience}, Hoboken, N.J, 2003.

\bibitem{kozin1988autoregressive}
F.~Kozin.
\newblock Autoregressive moving average models of earthquake records.
\newblock {\em Probabilistic Engineering Mechanics}, 3(2):58--63, 1988.

\bibitem{ma_identification_2005}
Y.~Ma and R.~Vidal.
\newblock Identification of deterministic switched {{ARX}} systems via
  identification of algebraic varieties.
\newblock In {\em International Workshop on Hybrid Systems: Computation and
  Control}, pages 449--465. {Springer}, 2005.

\bibitem{ogawa1993application}
T.~Ogawa, H.~Sonoda, S.~Ishiwa, and Y.~Shigeta.
\newblock An application of autoregressive model to pattern discrimination of
  brain electrical activity mapping.
\newblock {\em Brain topography}, 6(1):3--11, 1993.

\bibitem{ozay_set_2015}
N.~Ozay, C.~Lagoa, and M.~Sznaier.
\newblock Set membership identification of switched linear systems with known
  number of subsystems.
\newblock {\em Automatica}, 51:180--191, Jan. 2015.

\bibitem{ozay_sparsification_2012}
N.~Ozay, M.~Sznaier, C.~M. Lagoa, and O.~I. Camps.
\newblock A {{Sparsification Approach}} to {{Set Membership Identification}} of
  {{Switched Affine Systems}}.
\newblock {\em IEEE Trans. on Aut. Control}, 57(3):634--648, Mar. 2012.

\bibitem{strohmer2009randomized}
T.~Strohmer and R.~Vershynin.
\newblock A randomized kaczmarz algorithm with exponential convergence.
\newblock {\em Journal of Fourier Analysis and Applications}, 15(2):262--278,
  2009.

\bibitem{vidal_recursive_2008}
R.~Vidal.
\newblock Recursive identification of switched {{ARX}} systems.
\newblock {\em Automatica}, 44(9):2274--2287, Sept. 2008.

\bibitem{vidal_algebraic_2003}
R.~Vidal, S.~Soatto, Y.~Ma, and S.~Sastry.
\newblock An algebraic geometric approach to the identification of a class of
  linear hybrid systems.
\newblock In {\em Decision and Control, 2003. Proceedings. 42nd IEEE Conference
  on}, volume~1, pages 167--172. {IEEE}, 2003.

\end{thebibliography}

	\appendix
		
	
	\section{Proofs for Preliminary Results in Section \ref{subsctn_PreliminaryResult}}
	
	\subsection{{Proof for Lemma \ref{lemma_NoiseData}}}
	
	\begin{pf}
	Let $[N_R] = \{1, \dots, N_R \}$. From Assumption \ref{asmp_noise}, $\forall t, \expctn[n_t] = 0, \expctn[n_t^2] = \sigma_n^2$, and since $\forall i \in [N_R]$, $r_t(i)$ is a deterministic time step, so we have $\expctn[n_{r_t(i)}] = 0$ and $\expctn[n_{r_t(i)}^2] = \sigma_n^2$. Therefore,
	\begin{equation}
	\begin{split}
	\expctn[n_{r_t(l_t)}]
	&= \expctn[ \expctn [ n_{r_t(l_t)} | \Phi^R_{\wht{\sysidx}_t, t}, l_t=i]] \\
	&= \expctn[ \expctn [ n_{r_t(i)} | \Phi^R_{\wht{\sysidx}_t, t}, l_t=i]] \\
	&= \expctn[ \expctn [ n_{r_t(i)} | \Phi^R_{\wht{\sysidx}_t, t}]] \\
	&= \expctn[ n_{r_t(i)}] \\
	&= 0	
	\end{split}		
	\end{equation}
	where the third equality holds since given $\Phi^R_{\wht{\sysidx}_t, t}$, whether $l_t$ is chosen to be $i$ is independent of $n_{r_t(i)}$. Similarly,	
	\begin{equation}
	\begin{split}
	\expctn[n_{r_t(l_t)}^2]  
	&= \expctn[ \expctn [ n_{r_t(l_t)}^2 | \Phi^R_{\wht{\sysidx}_t, t}, l_t=i]] \\
	&= \expctn[ \expctn [ n_{r_t(i)}^2 | \Phi^R_{\wht{\sysidx}_t, t}, l_t=i]] \\
	&= \expctn[ \expctn [ n_{r_t(i)}^2 | \Phi^R_{\wht{\sysidx}_t, t}]] \\
	&= \expctn[ n_{r_t(i)}^2 ] \\
	& = \sigma_n^2
	\end{split}		
	\end{equation}
	So (\ref{lemmaresult_1}) is proved.
	
	From Assumption \ref{asmp_noise}, we know $\forall t, u_t$ and $n_t$ are independent, so $n_t$ is also independent of $\bvltn{\phi}_t$ from \eqref{eqn_ARXSystem}. Since $\forall i \in [N_R]$, $r_t(i)$ is a deterministic time step, we know $n_{r_t(i)}$ is independent of $\bvltn{\phi}_{r_t(i)}$. Therefore
	\begin{equation}
	\begin{split}
	\expctn[\bvltn{\phi}_{r_t(l_t)} n_{r_t(l_t)}]
	&= \expctn[\expctn[ \bvltn{\phi}_{r_t(l_t)} n_{r_t(l_t)} | \Phi^R_{\wht{\sysidx}_t, t}, l_t=i]] \\
	&= \expctn[\expctn[ \bvltn{\phi}_{r_t(i)} n_{r_t(i)} | \Phi^R_{\wht{\sysidx}_t, t}, l_t=i]] \\
	&= \expctn[\expctn[ \bvltn{\phi}_{r_t(i)} n_{r_t(i)} | \Phi^R_{\wht{\sysidx}_t, t}]] \\
	&= \expctn[ \bvltn{\phi}_{r_t(i)} n_{r_t(i)} ] \\
	&= \expctn[ \bvltn{\phi}_{r_t(i)} ] \expctn[ n_{r_t(i)} ] \\
	&= 0 \\
	&= \expctn[ \bvltn{\phi}_{r_t(l_t)}] \expctn[n_{r_t(l_t)}]
	\end{split}		
	\end{equation}	
	Therefore, $n_{r_t(l_t)}$ and $\bvltn{\phi}_{r_t(l_t)}$ are uncorrelated, and (\ref{lemmaresult_2}) is proved.
	
	From (\ref{lemmaresult_1}) and (\ref{lemmaresult_2}), we can see
	\begin{equation}
	\begin{split}
	\expctn \squarebracketsbig{\frac{n_{r_t(l_t)}^2}{\norm{\bvltn{\phi}_{r_t(l_t)}}^2}} 
	& = \expctn [n_{r_t(l_t)}^2] \expctn \squarebracketsbig{\frac{1}{\norm{\bvltn{\phi}_{r_t(l_t)}}^2}}\\
	& = \sigma_n^2 \expctn \squarebracketsbig{ \expctn \squarebracketsbig{ \left. \frac{1}{\norm{\bvltn{\phi}_{r_t(l_t)}}^2} \right| \Phi^R_{\wht{\sysidx}_t, t}  }   }\\
	& = \sigma_n^2 \expctn \squarebracketsbig{ \sum_{i\in [N_R]} \frac{1}{\norm{\bvltn{\phi}_{r_t(i)}}^2} P(l_t=i | \Phi^R_{\wht{\sysidx}_t, t} )  }\\
	& = \sigma_n^2 \expctn \squarebracketsbig{ \sum_{i\in [N_R]} \frac{1}{\norm{\bvltn{\phi}_{r_t(i)}}^2} \frac{\norm{\bvltn{\phi}_{r_t(i)}}^2}{\norm{\bvltn{\Phi}^R_{{\wht{\sysidx}_t}, t}}_F^2}  }\\
	& = \sigma_n^2 \expctn \squarebracketsbig{ \frac{N_R}{\norm{\bvltn{\Phi}^R_{{\wht{\sysidx}_t}, t}}_F^2}  }\\		
	& = \sigma_n^2 N_R \expctn \squarebracketsbig{ \frac{1}{\norm{\bvltn{\Phi}^R_{{\wht{\sysidx}_t}, t}}_F^2}  }\\	
	\end{split}
	\end{equation}
	From Lemma \ref{lemma_boundsonNorm}, we have $F_{\min} \leq \norm{\bvltn{\Phi}^R_{i,t}}_F \leq F_{\max}$, so (\ref{lemmaresult_3}) is proved.
	\end{pf}

	\subsection{{Proof for Lemma \ref{lemma_boundonExpectation}}}
	
	\begin{pf}
	First we prove the lower bound. Let ${\bvltn{\Phi}^R_{\wht{\sysidx}_t,t}}^{-1}$ denote the right inverse of $\bvltn{\Phi}^R_{\wht{\sysidx}_t,t}$, then accordingly  ${{\bvltn{\Phi}^R_{\wht{\sysidx}_t,t}}^{-1}}^\T$ is the left inverse of ${\bvltn{\Phi}^R_{\wht{\sysidx}_t,t}}^\T$. As for $\norm{{\bvltn{\Phi}^R_{\wht{\sysidx}_t,t}}^{-1}}_2$, by definition of matrix norm, we have, for $\forall \, \bv{z}$,
	\begin{equation}
	\norm{{\bvltn{\Phi}^R_{\wht{\sysidx}_t,t}}^{-1}}_2 = \norm{{{\bvltn{\Phi}^R_{\wht{\sysidx}_t,t}}^{-1}}^\T}_2 \geq \frac{\norm{{{\bvltn{\Phi}^R_{\wht{\sysidx}_t,t}}^{-1}}^\T {{\bvltn{\Phi}^R_{\wht{\sysidx}_t,t}}^\T} \bv{z}}}{\norm{{{\bvltn{\Phi}^R_{\wht{\sysidx}_t,t}}^\T} \bv{z}}}
	\end{equation}
	which gives
	\begin{equation}
	\norm{{{\bvltn{\Phi}^R_{\wht{\sysidx}_t,t}}^\T} \bv{z}}^2 \geq \frac{\norm{\bv{z}}^2}{\norm{{\bvltn{\Phi}^R_{\wht{\sysidx}_t,t}}^{-1}}_2^2}
	\end{equation}
	Expanding LHS and dividing both sides by $\norm{\bvltn{\Phi}^R_{\wht{\sysidx}_t,t}}_F^2$, we have
	\begin{equation}
	\sum_{i\in[N_R]} \frac{1}{\norm{\bvltn{\Phi}^R_{\wht{\sysidx}_t,t}}_F^2}\parenthesesbig{\bvltn{\phi}_{r_t(i)}^\T \bv{z}}^2 \geq \frac{\norm{\bv{z}}^2}{\norm{\bvltn{\Phi}^R_{\wht{\sysidx}_t,t}}_F^2 \norm{{\bvltn{\Phi}^R_{\wht{\sysidx}_t,t}}^{-1}}_2^2}
	\end{equation}
	Use the definition  $\kappa(\bvltn{\Phi}^R_{\wht{\sysidx}_t,t}) = \norm{\bvltn{\Phi}^R_{\wht{\sysidx}_t,t}}_F \norm{{\bvltn{\Phi}^R_{\wht{\sysidx}_t,t}}^{-1}}_2 $ in Lemma \ref{lemma_boundsonNorm}, then
	\begin{equation}
	\sum_{i\in[N_R]} \frac{\norm{\bvltn{\phi}_{r_t(i)}}^2}{\norm{\bvltn{\Phi}^R_{\wht{\sysidx}_t,t}}_F^2} \parenthesesbig{\frac{\bvltn{\phi}_{r_t(i)}^\T \bv{z}}{\norm{\bvltn{\phi}_{r_t(i)}} }}^2
	\geq
	\kappa(\bvltn{\Phi}^R_{\wht{\sysidx}_t,t})^{-2} \norm{\bv{z}}^2
	\end{equation}
	Note that the LHS is equal to $\expctn \squarebracketsbig{ \left. \parentheses{  \frac{\bvltn{\phi}_{r_t(l_t)}^\T \bv{z} }{\norm{\bvltn{\phi}_{r_t(l_t)}}}  }^2 \right| \bv{z}, \bvltn{\Phi}^R_{\wht{\sysidx}_t,t} }$, so
	\begin{equation} \label{eqn_untitle34}
	\expctn \squarebracketsbig{\left. \parenthesesbig{ \frac{\bvltn{\phi}_{r_t(l_t)}^\T \bv{z} }{\norm{\bvltn{\phi}_{r_t(l_t)}}}  }^2 \right| \bv{z}, \bvltn{\Phi}^R_{\wht{\sysidx}_t,t}} \geq \kappa(\bvltn{\Phi}^R_{\wht{\sysidx}_t,t})^{-2} \norm{\bv{z}}^2
	\end{equation}
	Now taking expectation of both sides again and using smoothing property of expectation, we have
	\begin{equation}
	\expctn \squarebracketsbig{\parenthesesbig{\frac{\bvltn{\phi}_{{r_t(l_t)}}^\T \bv{z} }{\norm{\bvltn{\phi}_{{r_t(l_t)}}}}  }^2 } \geq \expctn \squarebracketsbig{ \kappa(\bvltn{\Phi}^R_{\wht{\sysidx}_t,t})^{-2}  \norm{\bv{z}}^2 }
	\end{equation}	
	From Lemma \ref{lemma_boundsonNorm}, we have $\kappa(\bvltn{\Phi}^R_{i,t}) \leq \kappa_{\max}$, so
	\begin{equation}
	\expctn \squarebracketsbig{\parenthesesbig{\frac{\bvltn{\phi}_{{r_t(l_t)}}^\T \bv{z} }{\norm{\bvltn{\phi}_{{r_t(l_t)}}}}  }^2 } \geq \kappa_{\max}^{-2}  \expctn \squarebracketsbig{  \norm{\bv{z}}^2 }
	\end{equation}
	
	As for the upper bound, note that for $\forall \bv{z}$,
	\begin{equation}
	\norm{\bvltn{\Phi}^R_{\wht{\sysidx}_t,t}}_2 = \norm{{\bvltn{\Phi}^R_{\wht{\sysidx}_t,t}}^\T}_2 \geq \frac{\norm{{{\bvltn{\Phi}^R_{\wht{\sysidx}_t,t}}^\T} \bv{z}}}{\norm{\bv{z}}}
	\end{equation}
	which gives
	\begin{equation}
	\norm{{{\bvltn{\Phi}^R_{\wht{\sysidx}_t,t}}^\T} \bv{z}}^2 \leq {\norm{\bv{z}}}^2 \norm{\bvltn{\Phi}^R_{\wht{\sysidx}_t,t}}_2^2
	\end{equation}
	Then using similar technique as the proof for lower bound, we could have
	\begin{equation}  \label{eqn_untitle35}
	\expctn \squarebracketsbig{\parenthesesbig{\frac{\bvltn{\phi}_{{r_t(l_t)}}^\T \bv{z} }{\norm{\bvltn{\phi}_{{r_t(l_t)}}}}  }^2}
	\leq
	\xi_{\min}^{-2} \expctn[\norm{\bv{z}}^2]
	\end{equation}
	\end{pf}

			
	\section{Proofs for Valid Upper Bound Results in Section \ref{subsctn_ValidUpBd}}
	
	\subsection{{Proof for Theorem \ref{thrm_validUpperbound}}}
	
	\begin{pf}
	From the setup statement in Theorem \ref{thrm_validUpperbound},  we could see that to show the theorem, it suffices to consider there's only one subsystem, namely subsystem i, in the hybrid SARX model. Then, $c_i=t$, and the setup condition in theorem statement can be met automatically when $t \geq N_C$.
	
	When $t \geq N_C$, i.e. $c_i \geq N_C$: the data $\bvltn{\phi}_t^*, y_t^*, \eta_t^*$ we choose to update the candidate in Line \ref{algline_updateEstimate} of Algorithm \ref{alg_main} is formed in Line \ref{algline_untitled5}, where we sample a column index $l_t$ from the matrix $\bvltn{\Phi}^R_{i,t}$ in Line \ref{algline_PickColumn} of Algorithm \ref{alg_main}. Since $\bvltn{\Phi}^R_{i,t}$ is a matrix with columns being data vectors collected at different time, we essentially sampled a time index. Let $r_t(l_t)$ denote the true time index corresponding to the column $l_t$ we sample at time $t$. So $\bvltn{\phi}_t^* = \bvltn{\phi}_{r_t(l_t)}, y_t^* = y_{r_t(l_t)}$. In addition, we let $n_t^* = n_{r_t(l_t)}$. 
		
	Plugging definition $\bvltn{\epsilon}_{i, t} = \bv{w}_i - \wht{\bv{w}}_{i, t}$ and system equation $y_{t}^* = \bv{w}_i^\T \bvltn{\phi}_{t}^* + n_{t}^*$ into update rule $\wht{\bv{w}}_{t} = \wht{\bv{w}}_{t-1} - \eta_t^* \bvltn{\phi}_{t}^* \parentheses{\wht{\bv{w}}_{t-1}^\T  \bvltn{\phi}_{t}^* - y_t^* }$, we could have
	\begin{equation} \label{eqn_ErrDynamics}
	\begin{split}
	\bvltn{\epsilon}_{i, t} 
	&= (I - \eta_{t}^* \bvltn{\phi}_{t}^* {\bvltn{\phi}_{t}^*}^\T ) \bvltn{\epsilon}_{i, t-1} - \eta_{t}^* \bvltn{\phi}_{t}^* n_{t}^* \\
	&= \bvltn{\epsilon}_{i, t-1} - \eta_{t}^* \bvltn{\phi}_{t}^* {\bvltn{\phi}_{t}^*}^\T \bvltn{\epsilon}_{i, t-1} - \eta_{t}^* \bvltn{\phi}_{t}^* n_{t}^*
	\end{split}
	\end{equation}
	Replacing the first term $\bvltn{\epsilon}_{i, t-1}$ on the RHS of \eqref{eqn_ErrDynamics} by $\bvltn{\epsilon}_{i, t-1} =  \bvltn{\epsilon}_{i, t-2} - \eta_{t-1}^* \bvltn{\phi}_{t-1}^* {\bvltn{\phi}_{t-1}^*}^\T \bvltn{\epsilon}_{i, t-1} - \eta_{t-1}^* \bvltn{\phi}_{t-1}^* n_{t-1}^* $ and repeat this procedure recursively, we could finally have
	\begin{equation} \label{eqn_z_a}
	\bvltn{\epsilon}_{i, t}  
	= \bvltn{\epsilon}_{i, t - N_C} 
	- \sum_{j=t}^{t-(N_C-1)} \eta_{j}^* \bvltn{\phi}_{j}^* {\bvltn{\phi}_{j}^*}^\T \bvltn{\epsilon}_{i, j-1} 
	- \sum_{j=t}^{t-(N_C-1)} \eta_{j}^* \bvltn{\phi}_{j}^* n_{j}^*
	\end{equation}
	Consider the LHS of \eqref{eqn_z_a}, by addition and subtraction, we could see
	\begin{equation} \label{eqn_z_b}
	\bvltn{\epsilon}_{i, t}  
	= \sum_{j=t}^{t-(N_C-1)} \eta_{j}^* \bvltn{\phi}_{j}^* {\bvltn{\phi}_{j}^*}^\T \bvltn{\epsilon}_{i, t}
	+ \bvltn{\epsilon}_{i, t}  
	- \sum_{j=t}^{t-(N_C-1)} \eta_{j}^* \bvltn{\phi}_{j}^* {\bvltn{\phi}_{j}^*}^\T \bvltn{\epsilon}_{i, t}
	\end{equation}										
	Combining \eqref{eqn_z_a} and \eqref{eqn_z_b}, we have 
	\begin{equation}
	\sum_{j=t}^{t-(N_C-1)} \eta_{j}^* \bvltn{\phi}_{j}^* {\bvltn{\phi}_{j}^*}^\T \bvltn{\epsilon}_{i, t} = 
	\parenthesesbig{
	\parentheses{\wht{\bv{w}}_{i,t} - \wht{\bv{w}}_{i,t-N_C}}
	- \sum_{j=t}^{t-(N_C-1)} \eta_{j}^* \bvltn{\phi}_{j}^* {\bvltn{\phi}_{j}^*}^\T \parentheses{\wht{\bv{w}}_{i,t} - \wht{\bv{w}}_{i,j-1}}
	}  \\
	- \sum_{j=t}^{t-(N_C-1)} \eta_{j}^* \bvltn{\phi}_{j}^* n_{j}^*
	\end{equation}
	Now using the notations and operator defined in Line \ref{algline_untitled6} to Line \ref{algline_untitled7} in Algorithm \ref{alg_computeUpperbound}, and let $\splitatcommas{\bv{n}_t = [n_{t-(N_C-1)}, \dots,  n_{t-1}, n_{t}]^\T }$ we have a neat form:
	\begin{equation} \label{eqn_untitled6}
	\bvltn{\Phi}^C_{i,t} \bv{H} {\bvltn{\Phi}^C_{i,t}}^\T \bvltn{\epsilon}_{i, t}
	=
	\squarebracketsbig{\Delta \wht{\bv{w}} - \bvltn{\Phi}^C_{i,t} \square \parenthesesbig{\bvltn{\Phi}^C_{i,t}, \Delta \wht{\bv{W}}}} - \bvltn{\Phi}^C_{i,t} \bv{H} \bv{n}_t
	\end{equation}
	
	Now, we want to show the invertibility of matrix $\bvltn{\Phi}^C_{i,t} \bv{H} {\bvltn{\Phi}^C_{i,t}}^\T $. Define
	\begin{equation}
	\wtd{\bvltn{\Phi}} {=} \squarebracketsbig{\sqrt{\eta_{ t-(N_C-1) }^* } \bvltn{\phi}_{ t-(N_C-1) }^* , \dots,  \sqrt{\eta_{ t-1 }^* } \bvltn{\phi}_{ t-1 }^*, \sqrt{\eta_{ t }^* } \bvltn{\phi}_{ t }^*}_{n \x N_C}
	\end{equation}
	, then we could see that
	\begin{equation}
	\bvltn{\Phi}^C_{i,t} \bv{H} {\bvltn{\Phi}^C_{i,t}}^\T = \sum_{j=t}^{t-(N_C-1)} \eta_{j}^* \bvltn{\phi}_{j}^* {\bvltn{\phi}_{j}^*}^\T = \wtd{\bvltn{\Phi}} \wtd{\bvltn{\Phi}}^\T
	\end{equation}
	, so it suffices to show $\wtd{\bvltn{\Phi}}$ has $n$ linearly independent columns.	Note that $\eta^* >0$ and $\splitatcommas{\bvltn{\Phi}^C_{i,t} = [\bvltn{\phi}_{t-(N_C-1)}^* , \dots,   \bvltn{\phi}_{t-1}^*,  \bvltn{\phi}_{t}^*] }$, so it further suffices to show $\bvltn{\Phi}^C_{i,t}$ has $n$ linearly independent columns.
	
	Since $\bvltn{\Phi}^C_{i,t}$ is composed of $N_C$ columns sample from different matrices $\bvltn{\Phi}^R_{i,t} \in \R^{n \times N_R}$ from time $t-(N_C-1)$ to $t$, then the condition $N_C \geq N_R^2$ requirement in Algorithm \ref{alg_main} guarantees that there are at least $N_R$ columns in $\bvltn{\Phi}^C_{i,t}$ such that their generating time are different. Then from Assumption \ref{asmp_singularvalue}, we know these $n$ columns must be linearly independent, and so are the corresponding columns in $\wtd{\bvltn{\Phi}}$. Therefore, $\bvltn{\Phi}^C_{i,t} \bv{H} {\bvltn{\Phi}^C_{i,t}}^\T$ is invertible.
	
	With this result, \eqref{eqn_untitled6} becomes:
	\begin{equation} \label{eqn_z_d}
	\bvltn{\epsilon}_{i, t}
	=
	\parentheses{\bvltn{\Phi}^C_{i,t} \bv{H} {\bvltn{\Phi}^C_{i,t}}^\T}^{-1} \squarebracketsbig{\Delta \wht{\bv{w}} - \bvltn{\Phi}^C_{i,t} \square \parenthesesbig{\bvltn{\Phi}^C_{i,t}, \Delta \wht{\bv{W}}}} \\
	- \parentheses{\bvltn{\Phi}^C_{i,t} \bv{H} {\bvltn{\Phi}^C_{i,t}}^\T}^{-1}  \bvltn{\Phi}^C_{i,t} \bv{H} \bv{n}_t
	\end{equation}	
	then using definition of $\bv{A}$ and $\bv{b}$ in Algorithm \ref{alg_computeUpperbound}, we have
	\begin{equation} \label{eqn_untitle45}
	\bvltn{\epsilon}_{i,t} = \bv{b} - \bv{A} \bv{n}_t
	\end{equation}
	Since $\norm{\bv{n}_t}_\infty \leq n_{\max}$, if we define the set of vertices $V = \{[\pm n_{\max}, \pm n_{\max}, \dots,  \pm n_{\max}]_{N_C}^\T \}$, then it's easy to see
	\begin{equation}
	\norm{\bv{b} - \bv{A} \bv{n}_t} \leq \max_{\bv{n}\in V} \norm{A \bv{n} - \bv{b}}
	\end{equation}
	And since $\epsilon_{i, t}^u \equiv \max_{\bv{n}\in V} \norm{A \bv{n} - \bv{b}}$, we could finally see
	\begin{equation}
	\epsilon_{i, t}^u \geq \norm{\bvltn{\epsilon}_{i,t}}
	\end{equation}
	\end{pf}

	
	\section{Proofs for Partial Convergence in Section \ref{subsctn_PartialConvrg}}
	
	\subsection{{Proof for Lemma \ref{lemma_InitPhasePartial}}}
	
	\begin{pf}
	According Algorithm \ref{alg_main}, when $t\leq N_R-1$, we know $c_i < N_R$, and the update rule is given by $\wht{\bv{w}}_{i,t} = \wht{\bv{w}}_{i,t-1} - \eta_t^* \bvltn{\phi}_t (\wht{\bv{w}}_{i,t}^\T \bvltn{\phi}_t - y_t)$. Since $\bvltn{\epsilon}_{i,t} = \bv{w}_i - \wht{\bv{w}}_{i,t}$ and $\bv{w}_i^\T \bvltn{\phi}_{t} + n_{t} = y_{t}$, we can derive the following error dynamics through simple algebra:
	\begin{equation} \label{eqn_untitled17}
	\bvltn{\epsilon}_{i,t} = (I - \eta_t^* \bvltn{\phi}_{t} \bvltn{\phi}_{t}^\T) \bvltn{\epsilon}_{i, t-1} - \eta_t^* \bvltn{\phi}_{t} n_{t}
	\end{equation}	
	Notice that in Algorithm \ref{alg_main}, we set $\eta_t^* = \norm{\bvltn{\phi}_{t}}^{-2}$, so
	\begin{equation}
	\bvltn{\epsilon}_{i,t} = \parenthesesbig{I - \frac{\bvltn{\phi}_{t} \bvltn{\phi}_{t}^\T}{\norm{\bvltn{\phi}_{t}}^{2}}} \bvltn{\epsilon}_{i, t-1} - \frac{\bvltn{\phi}_{t} n_{t}}{\norm{\bvltn{\phi}_{t}}^{2}} 
	\end{equation}
	Taking norm squares of both sides,
	\begin{equation} \label{eqn_untitle18}
	\norm{\bvltn{\epsilon}_{i,t}}^2 = 
	\normbig{\parenthesesbig{I - \frac{\bvltn{\phi}_{t} \bvltn{\phi}_{t}^\T}{\norm{\bvltn{\phi}_{t}}^{2}}} \frac{\bvltn{\epsilon}_{i, t-1}}{\norm{\bvltn{\epsilon}_{i, t-1}}}}^2 \norm{\bvltn{\epsilon}_{i, t-1}}^2
	+ \frac{n_{t}^2}{\norm{\bvltn{\phi}_{t}}^{2}}
	\end{equation}
	of which the cross term vanishes because it's equal to 0. Now consider the first term in \eqref{eqn_untitle18},
	\begin{equation} \label{eqn_untitle19}
	\begin{split}
	&\normbig{\parenthesesbig{I - \frac{\bvltn{\phi}_{t} \bvltn{\phi}_{t}^\T}{\norm{\bvltn{\phi}_{t}}^{2}}} \frac{\bvltn{\epsilon}_{i, t-1}}{\norm{\bvltn{\epsilon}_{i, t-1}}}}^2 \\
	=& \frac{\bvltn{\epsilon}_{i, t-1}^\T}{\norm{\bvltn{\epsilon}_{i, t-1}}} \parenthesesbig{I - \frac{\bvltn{\phi}_{t} \bvltn{\phi}_{t}^\T}{\norm{\bvltn{\phi}_{t}}^{2}}} \frac{\bvltn{\epsilon}_{i, t-1}}{\norm{\bvltn{\epsilon}_{i, t-1}}} \\
	=& 1-\parenthesesbig{\frac{\bvltn{\phi}_{t}^\T \bvltn{\epsilon}_{i, t-1}}{ \norm{\bvltn{\phi}_{t}} \norm{\bvltn{\epsilon}_{i, t-1}} } }^2
	\end{split}		
	\end{equation}
	Plugging \eqref{eqn_untitle19} into \eqref{eqn_untitle18}, then
	\begin{equation}
	\norm{\bvltn{\epsilon}_{i,t}}^2 = 
	\squarebracketsbig{1-\parenthesesbig{\frac{\bvltn{\phi}_{t}^\T \bvltn{\epsilon}_{i, t-1}}{ \norm{\bvltn{\phi}_{t}} \norm{\bvltn{\epsilon}_{i, t-1}} } }^2} \norm{\bvltn{\epsilon}_{i, t-1}}^2
	+ \frac{n_{t}^2}{\norm{\bvltn{\phi}_{t}}^{2}}
	\end{equation}
	Since $0 \leq \parentheses{\frac{\bvltn{\phi}_{t}^\T \bvltn{\epsilon}_{i, t-1}}{ \norm{\bvltn{\phi}_{t}} \norm{\bvltn{\epsilon}_{i, t-1}} } }^2 \leq 1$, we have
	\begin{equation} \label{eqn_untitle20}
	\frac{n_{t}^2}{\norm{\bvltn{\phi}_{t}}^{2}} 
	\leq 
	\norm{\bvltn{\epsilon}_{i,t}}^2 
	\leq 
	\norm{\bvltn{\epsilon}_{i, t-1}}^2
	+ \frac{n_{t}^2}{\norm{\bvltn{\phi}_{t}}^{2}}
	\end{equation}
	Since $\norm{\bvltn{\phi}_t}^2 \leq \phi_{\max}^2$ and $\frac{\norm{\bvltn{\phi}_t}}{|n_t|} \geq S_{\min}$ according to Assumption \ref{asmp_data}, then
	\begin{equation} \label{eqn_untitle21}
	\frac{n_{t}^2}{\phi_{\max}^2}
	\leq 
	\norm{\bvltn{\epsilon}_{i,t}}^2 
	\leq 
	\norm{\bvltn{\epsilon}_{i, t-1}}^2
	+ \frac{1}{S_{\min}^2}
	\end{equation}
	Now taking expectation of both sides of \eqref{eqn_untitle21},
	\begin{equation} \label{eqn_untitle25}
	\frac{\sigma_n^2}{\phi_{\max}^2}
	\leq 
	\expctn \squarebracketsbig{\norm{\bvltn{\epsilon}_{i,t}}^2 }
	\leq 
	\expctn \squarebracketsbig{ \norm{\bvltn{\epsilon}_{i, t-1}}^2 }
	+ \frac{1}{S_{\min}^2}
	\end{equation}	
	Now if we apply \eqref{eqn_untitle25} recursively, we could finally prove \eqref{eqn_InitPhasePartial} in the lemma.
	\end{pf}
	
	\subsection{{Proof for Lemma \ref{lemma_2ndConvergenceConstant}}}
	
	\begin{pf}
	From Algorithm \ref{alg_main}, when $t\geq N_R$, $c_i \geq N_R$. And to update estimate, we first sample a column index $l_t$ from the matrix $\bvltn{\Phi}^R_{i,t}$ in Line \ref{algline_PickColumn} of Algorithm \ref{alg_main}. Since $\bvltn{\Phi}^R_{i,t}$ is a matrix with columns being data vectors collected at different time, we essentially sampled a time index. Let $r_t(l_t)$ denote the true time index corresponding to the column $l_t$ we sample at time $t$. So the corresponding $\bvltn{\phi}_t^*, y_t^*$ are actually $\bvltn{\phi}_{r_{t(l)}}, y_{r_t(l_t)}$. And we have the update rule $\wht{\bv{w}}_{i,t} = \wht{\bv{w}}_{i,t-1} - \eta_t^* \bvltn{\phi}_{r_t(l_t)} (\wht{\bv{w}}_{t-1}^\T \bvltn{\phi}_{r_t(l_t)} - y_{r_t(l_t)})$. So following \eqref{eqn_untitle18} in proof for Lemma \ref{lemma_InitPhasePartial}, we have
	\begin{equation} \label{eqn_untitle17}
	\norm{\bvltn{\epsilon}_{i,t}}^2 = 
	\squarebracketsbig{1-\parenthesesbig{\frac{\bvltn{\phi}_{r_t(l_t)}^\T \bvltn{\epsilon}_{i,t-1}}{ \norm{\bvltn{\phi}_{r_t(l_t)}} \norm{\bvltn{\epsilon}_{i,t-1}} } }^2} \norm{\bvltn{\epsilon}_{i,t-1}}^2
	+ \frac{n_{r_t(l_t)}^2}{\norm{\bvltn{\phi}_{r_t(l_t)}}^{2}}
	\end{equation}
	Now take expectation of both sides of \eqref{eqn_untitle17},
	\begin{equation} \label{eqn_untitle26}
	\expctn[\norm{\bvltn{\epsilon}_{i,t}}^2 ] = 
	\expctn \squarebracketsbig{\norm{\bvltn{\epsilon}_{i,t-1}}^2} 
	- \expctn \squarebracketsbig{\parenthesesbig{\frac{\bvltn{\phi}_{r_t(l_t)}^\T \bvltn{\epsilon}_{i,t-1}}{ \norm{\bvltn{\phi}_{r_t(l_t)}}  } }^2  } 
	+ \expctn \squarebracketsbig{ \frac{n_{r_t(l_t)}^2}{\norm{\bvltn{\phi}_{r_t(l_t)}}^{2}} }
	\end{equation}
	Applying Lemma \ref{lemma_boundonExpectation} and Lemma \ref{lemma_NoiseData}(\ref{lemmaresult_3}), we have
	\begin{equation} \label{eqn_untitle30}
	\left\{\begin{matrix}
	\expctn \squarebracketsbig{\norm{\bvltn{\epsilon}_{i,t}}^2 }
	{\geq}
	\parenthesesbig{1-\xi_{\min}^{-2}} \expctn \squarebracketsbig{\norm{\bvltn{\epsilon}_{i,t-1}}^2} + \frac{N_R}{F_{\max}^2}\sigma_n^2	\\ 
	\expctn \squarebracketsbig{\norm{\bvltn{\epsilon}_{i,t}}^2 }
	{\leq}
	\parenthesesbig{1-\kappa_{\max}^{-2}} \expctn \squarebracketsbig{\norm{\bvltn{\epsilon}_{i,t-1}}^2} + \frac{N_R}{F_{\min}^2}\sigma_n^2
	
	\end{matrix}\right.
	\end{equation}

	Finally, apply \eqref{eqn_untitle30} recursively, we could end up getting \eqref{eqn_2ndConvergenceConstant_Lower} and \eqref{eqn_2ndConvergenceConstant_Upper} in Lemma \ref{lemma_2ndConvergenceConstant}
	\end{pf}
	
	\subsection{{Proof for Theorem \ref{thrm_PartialConstant}}}
	
	\begin{pf}
	When there is only one subsystem, Lemma \ref{lemma_InitPhasePartial} and Lemma \ref{lemma_2ndConvergenceConstant} selectively characterize the behavior of estimation error when $t<N_R$ and $t \geq N_R$. By combining them and replacing the universal time index $t$ in Lemma \ref{lemma_InitPhasePartial} and Lemma \ref{lemma_2ndConvergenceConstant} with the individual time index $r(i,t)$ for subsystem $i$, we can have this theorem.	
	\end{pf}

			
	\section{Proofs for Local Convergence in Section \ref{subsctn_LocalConvrg}}
	
	\subsection{{Proof for Lemma \ref{lemma_Supermartingale}}}
	
	\begin{pf}	
	Let $\bvltn{\rchi}_t$ be ``All data $\curlybrackets{\bvltn{\phi}, y}$ assigned to candidate i up to time t and all the data $\curlybrackets{\bvltn{\phi}^*, y^*}$ we used to update $\wht{\bv{w}}_i$  up to time $t$''. Then we can see $\bvltn{\rchi}_t \subset \bvltn{\rchi}_{t+1}$ and 
	\begin{equation} \label{eqn_untitle31}
	\expctn \squarebracketsbig{\left. \norm{\bvltn{\epsilon}_{i, t}}^2 \right| \bvltn{\rchi}_t} 
	= \expctn \squarebracketsbig{\left. \norm{\bv{w}_i - \wht{\bv{w}}_{i,t}}^2 \right| \bvltn{\rchi}_t} 
	= \norm{\bv{w}_i - \wht{\bv{w}}_{i,t}}^2
	= \norm{\bvltn{\epsilon}_{i, t}}^2
	\end{equation}
	where the second equality holds since knowing $\bvltn{\rchi}_t$ we know update process of $\wht{\bv{w}}_{i,0}, \wht{\bv{w}}_{i,1}, \dots, \wht{\bv{w}}_{i,t}$ completely. \eqref{eqn_untitle31} says the randomness of $\norm{\bvltn{\epsilon}_{i, t}}^2$ completely comes from $\bvltn{\rchi}_t$.
	
	When $t \geq N_R$, \eqref{eqn_untitle17} characterizes the error dynamics, and we restate it here:
	\begin{equation} \label{eqn_untitle32}
	\norm{\bvltn{\epsilon}_{i,t}}^2 = 
	\norm{\bvltn{\epsilon}_{i,t-1}}^2 -\parenthesesbig{\frac{\bvltn{\phi}_{r_t(l_t)}^\T \bvltn{\epsilon}_{i,t-1}}{ \norm{\bvltn{\phi}_{r_t(l_t)}} } }^2
	+ \frac{n_{r_t(l_t)}^2}{\norm{\bvltn{\phi}_{r_t(l_t)}}^{2}}
	\end{equation}
	Now take $\expctn[\cdot | \bvltn{\rchi}_{t-1}]$ on both sides of \eqref{eqn_untitle32}, we have
	\begin{equation} \label{eqn_untitle33}
	\expctn \squarebracketsbig{ \norm{\bvltn{\epsilon}_{i,t}}^2 | \bvltn{\rchi}_{t-1} }
	{=} 
	\norm{\bvltn{\epsilon}_{i,t-1}}^2 
	{-} 
	\expctn \squarebracketsbig{ \left. \parenthesesbig{\frac{\bvltn{\phi}_{r_t(l_t)}^\T \bvltn{\epsilon}_{i,t-1}}{ \norm{\bvltn{\phi}_{r_t(l_t)}} } }^2 \right| \bvltn{\rchi}_{t-1} } 
	+ \expctn \squarebracketsbig{ \left. \frac{n_{r_t(l_t)}^2}{\norm{\bvltn{\phi}_{r_t(l_t)}}^{2}} \right| \bvltn{\rchi}_{t-1} }
	\end{equation}		
	First consider $\expctn \squarebracketsbig{ \left. \parenthesesbig{\frac{\bvltn{\phi}_{r_t(l_t)}^\T \bvltn{\epsilon}_{i,t-1}}{ \norm{\bvltn{\phi}_{r_t(l_t)}} } }^2 \right| \bvltn{\rchi}_{t-1} }$, we have
	\begin{equation} \label{eqn_untitle53}
	\begin{split}
	&\expctn \squarebracketsbig{ \left. \parenthesesbig{\frac{\bvltn{\phi}_{r_t(l_t)}^\T \bvltn{\epsilon}_{i,t-1}}{ \norm{\bvltn{\phi}_{r_t(l_t)}} } }^2 \right| \bvltn{\rchi}_{t-1} } \\
	=& \expctn \squarebracketsbig{ \left. \parenthesesbig{\frac{\bvltn{\phi}_{r_t(l_t)}^\T \bvltn{\epsilon}_{i,t-1}}{ \norm{\bvltn{\phi}_{r_t(l_t)}} } }^2 \right| \bvltn{\rchi}_{t-1}, \bvltn{\epsilon}_{i,t-1}} \\
	=& \expctn \squarebracketsbig{ \left. \parenthesesbig{\frac{\bvltn{\phi}_{r_t(l_t)}^\T \bvltn{\epsilon}_{i,t-1}}{ \norm{\bvltn{\phi}_{r_t(l_t)}} } }^2 \right| \bvltn{\Phi}^R_{i, t-1}, \bvltn{\epsilon}_{i,t-1}} \\		
	\end{split}
	\end{equation}
	where the first equality holds as $\bvltn{\epsilon}_{i, t-1}$ is nonrandom given $\bvltn{\rchi}_{t-1}$; the second equality holds for the following reason: $\bvltn{\Phi}^R_{i, t-1}$ can be determined from $\bvltn{\rchi}_{t-1}$, and $\bvltn{\phi}_{r_t(l_t)}$ is drawn from $\bvltn{\Phi}^R_{i, t-1}$ in an independent experiment, so $\bvltn{\phi}_{r_t(l_t)}$ depends on $\bvltn{\rchi}_{t-1}$ only through $\bvltn{\Phi}^R_{i, t-1}$. Note that RHS in \eqref{eqn_untitle53} can follow similar argument from \eqref{eqn_untitle34} to \eqref{eqn_untitle35}, then
	\begin{equation} \label{eqn_untitle37}
	\kappa_{\max}^{-2} \norm{\bvltn{\epsilon}_{i,t-1}}^2
	\leq 
	\expctn \squarebracketsbig{ \left. \parenthesesbig{\frac{\bvltn{\phi}_{r_t(l_t)}^\T \bvltn{\epsilon}_{i,t-1}}{ \norm{\bvltn{\phi}_{r_t(l_t)}} } }^2 \right| \bvltn{\rchi}_{t-1} } 
	\leq 
	\xi_{\min}^{-2} \norm{\bvltn{\epsilon}_{i,t-1}}^2 
	\end{equation}	
	Then consider $\expctn \squarebracketsbig{ \left. \frac{n_{r_t(l_t)}^2}{\norm{\bvltn{\phi}_{r_t(l_t)}}^{2}} \right| \bvltn{\rchi}_{t-1} }$. By Assumption \ref{asmp_data}, we have
	\begin{equation} \label{eqn_untitle36}
	\expctn \squarebracketsbig{ \left. \frac{n_{r_t(l_t)}^2}{\norm{\bvltn{\phi}_{r_t(l_t)}}^{2}} \right| \bvltn{\rchi}_{t-1} }
	\leq
	\frac{1}{S_{\min}^2}
	\end{equation}
	Applying \eqref{eqn_untitle36} and \eqref{eqn_untitle37} to \eqref{eqn_untitle33}, we have
	\begin{equation} \label{eqn_untitle39}
	\expctn \squarebracketsbig{ \norm{\bvltn{\epsilon}_{i,t}}^2 | \bvltn{\rchi}_{t-1} }
	\leq
	\parenthesesbig{1-\kappa_{\max}^{-2}} \norm{\bvltn{\epsilon}_{i,t-1}}^2 + \frac{1}{S_{\min}^2}
	\end{equation}
	
	Now we want to show $\expctn \squarebracketsbig{ \norm{\bvltn{\epsilon}_{i,t}}^2 | \bvltn{\rchi}_{t-1} } \leq \norm{\bvltn{\epsilon}_{i,t-1}}^2$. The general form of \eqref{eqn_untitle32} for $\forall t$ is 
	\begin{equation}
	\norm{\bvltn{\epsilon}_{i,t}}^2 = 
	\squarebracketsbig{1-\parenthesesbig{\frac{{\bvltn{\phi}_{t}^*}^\T \bvltn{\epsilon}_{i,t-1}}{ \norm{{\bvltn{\phi}_{t}^*}} \norm{\bvltn{\epsilon}_{i,t-1}} } }^2} \norm{\bvltn{\epsilon}_{i,t-1}}^2
	+ \frac{{n_{t}^*}^2}{\norm{{\bvltn{\phi}_{t}^*}}^{2}}
	\end{equation}
	where $\bvltn{\phi}_{t}^*$ is defined in Line \ref{algline_untitled4} and \ref{algline_untitled5} in Algorithm \ref{alg_main}, and we let $n_{t}^*$ denote the noise corresponding to data $\curlybrackets{\bvltn{\phi}_{t}^*, y_t^*}$. Since $\parenthesesbig{\frac{{\bvltn{\phi}_{t}^*}^\T \bvltn{\epsilon}_{i,t-1}}{ \norm{{\bvltn{\phi}_{t}^*}} \norm{\bvltn{\epsilon}_{i,t-1}} } }^2 {\leq} 1$ and by Assumption \ref{asmp_SNRUpperBound}, then $\forall t$
	\begin{equation}
	\norm{\bvltn{\epsilon}_{i,t}}^2 \geq \frac{{n_{t}^*}^2}{\norm{{\bvltn{\phi}_{t}^*}}^{2}} \geq \frac{1}{S_{\max}^2} \geq \frac{1}{\kappa_{\max}^2 S_{\min}^2}
	\end{equation}
	So for $\forall t \geq 2 $,
	\begin{equation} \label{eqn_untitle38}
	\norm{\bvltn{\epsilon}_{i,t-1}}^2  \geq \frac{1}{\kappa_{\max}^2 S_{\min}^2}
	\end{equation}
	Following \eqref{eqn_untitle38}, \eqref{eqn_untitle39} gives
	\begin{equation}
	\expctn \squarebracketsbig{ \norm{\bvltn{\epsilon}_{i,t}}^2 | \bvltn{\rchi}_{t-1} } \leq \norm{\bvltn{\epsilon}_{i,t-1}}^2
	\end{equation}
	So, we can see $\curlybrackets{\norm{\bvltn{\epsilon}_{i,t}}^2, t \geq N_R-1}$ is a supermartingale with respect to $\curlybrackets{\bvltn{\rchi}_{t}, t\geq N_R-1}$. Finally, using supermartingale maxima inequality, we have \eqref{eqn_supermartingaleMaximaIneq} directly.
	\end{pf}
	
	\subsection{{Proof for Lemma \ref{lemma_ProbBoundforSingleSys}}}
	
	\begin{pf}
	In the first phase of the algorithm when $t \leq N_R-1$, using \eqref{eqn_untitle25} recursively, we have
	\begin{equation}
	\expctn \squarebracketsbig{\norm{\bvltn{\epsilon}_{i, t}}^2} 
	\leq
	\norm{\bvltn{\epsilon}_{i,0}}^2 + \frac{t}{S_{\min}^2}
	\leq
	\epsilon_0^2 + \frac{N_R}{S_{\min}^2}
	\end{equation}
	Define $\epsilon_a$ such that $\epsilon_a^2 = \sqrt{\epsilon'^2 N_R \parenthesesbig{\epsilon_0^2 + \frac{N_R}{S_{\min}^2}}} $, then from the condition in the statement of Lemma \ref{lemma_ProbBoundforSingleSys}, we can see $\epsilon_a^2 \leq \epsilon'^2$. According to Markov inequality, we have
	\begin{equation}
	P\parenthesesbig{\norm{\bvltn{\epsilon}_{i, t}}^2 \leq \epsilon_a^2} \geq 1- \frac{1}{\epsilon_a^2} \parenthesesbig{\epsilon_0^2 + \frac{N_R}{S_{\min}^2}}
	\end{equation}
	Then using union bound, we have
	\begin{equation} \label{eqn_untitle40}
	P\parenthesesbig{\bigcap_{\tau=1}^{N_R-1} \curlybracketsbig{\norm{\bvltn{\epsilon}_{i, \tau}}^2 \leq \epsilon_a^2} } 
	\geq 
	1- \frac{N_R}{\epsilon_a^2} \parenthesesbig{\epsilon_0^2 + \frac{N_R}{S_{\min}^2}}
	\end{equation}
	
	Now for $t \geq N_R$, we have
	\begin{equation}
	\begin{split}
	&P\parenthesesbig{\bigcap_{\tau=1}^{t} \curlybracketsbig{\norm{\bvltn{\epsilon}_{i, \tau}}^2 \leq \epsilon'^2} } \\
	\geq& P\parenthesesbig{\bigcap_{\tau=1}^{N_R-1} \curlybracketsbig{\norm{\bvltn{\epsilon}_{i, \tau}}^2 \leq \epsilon_a^2}, \bigcap_{\tau=N_R}^{t} \curlybracketsbig{\norm{\bvltn{\epsilon}_{i, \tau}}^2 \leq \epsilon'^2} } \\
	=& P\parenthesesbig{\bigcap_{\tau=1}^{N_R-1} \curlybracketsbig{\norm{\bvltn{\epsilon}_{i, \tau}}^2 \leq \epsilon_a^2}} \cdot 
	P\left( \left. \bigcap_{\tau=N_R}^{t} \curlybracketsbig{\norm{\bvltn{\epsilon}_{i, \tau}}^2 \leq \epsilon'^2} \right. \right.\left. \left| \norm{\bvltn{\epsilon}_{i, N_R-1}}^2 \leq \epsilon_a^2, \bigcap_{\tau=1}^{N_R-2} \curlybracketsbig{\norm{\bvltn{\epsilon}_{i, \tau}}^2 \leq \epsilon_a^2} \right. \right) \\	
	\geq& \squarebracketsbig{1- \frac{N_R}{\epsilon_a^2} \parenthesesbig{\epsilon_0^2 + \frac{N_R}{S_{\min}^2}}}
	\squarebracketsbig{1 - \frac{\epsilon_a^2}{\epsilon'^2}} \\
	\geq& 1- \frac{N_R}{\epsilon_a^2} \parenthesesbig{\epsilon_0^2 + \frac{N_R}{S_{\min}^2}} - \frac{\epsilon_a^2}{\epsilon'^2} \\
	=& 1- 2 \sqrt{ \frac{N_R}{\epsilon'^2} \parenthesesbig{\epsilon_0^2 + \frac{N_R}{S_{\min}^2}} }
	\end{split}
	\end{equation}
	where the first inequality holds since $\epsilon_a^2 \leq \epsilon'^2$; the third inequality holds by applying \eqref{eqn_untitle40} and \eqref{eqn_supermartingaleMaximaIneq} in Lemma \ref{lemma_Supermartingale}; the last line holds by plugging in the definition of $\epsilon_a$.
	\end{pf}
	
	\subsection{{Proof for Lemma \ref{lemma_noMisassignment}}}
	
	\begin{pf}
		In Line \ref{algline_makeAssignment} of Algorithm \ref{alg_main}, we make assignment according to
		\begin{equation}
		\wht{\sysidx}_t = \arg \min_i \ r_i \cdot \max\parenthesesbig{1, \alpha  \frac{ \norm{\wtd{\bv{w}}_{i,t} - \wht{\bv{w}}_{i,t-1}}}{2 (\epsilon_{i,t-1}^u + \nu )}}^\beta
		\end{equation}
		So, if data $\curlybrackets{\bvltn{\phi}_t, y_t}$ is generated by subsystem $i$, i.e. ${\sysidx}_t = i$, and we want it to be assigned to candidate $i$ according to Line \ref{algline_untitled2} in Algorithm \ref{alg_main}, it suffices to have $\forall j \neq i$
		\begin{equation} \label{eqn_untitled3}
		\ r_j \cdot \max\parenthesesbig{1, \alpha  \frac{ \norm{\wtd{\bv{w}}_{j,t} - \wht{\bv{w}}_{j,t-1}}}{2 (\epsilon_{j,t-1}^u + \nu )}}^\beta
		>
		\ r_i \cdot \max\parenthesesbig{1, \alpha  \frac{ \norm{\wtd{\bv{w}}_{i,t} - \wht{\bv{w}}_{i,t-1}}}{2 (\epsilon_{i,t-1}^u + \nu )}}^\beta
		\end{equation}		
		From Line \ref{algline_untitled2} in Algorithm \ref{alg_main}, we can see $\norm{\wtd{\bv{w}}_{i,t} - \wht{\bv{w}}_{i,t-1}} = \norm{\bvltn{\phi}_t}^{-1} | \wht{\bv{w}}_{i,t-1}^\T  \bvltn{\phi}_t - y_t |$. So, \eqref{eqn_untitled3} is equivalent to
		\begin{equation} \label{eqn_untitled4}
		\ r_j \cdot \max\parenthesesbig{1, \alpha  \frac{ \norm{\bvltn{\phi}_t}^{-1} | \wht{\bv{w}}_{j,t-1}^\T  \bvltn{\phi}_t - y_t |}{2 (\epsilon_{j,t-1}^u + \nu )}}^\beta
		> 
		\ r_i \cdot \max\parenthesesbig{1, \alpha  \frac{ \norm{\bvltn{\phi}_t}^{-1} | \wht{\bv{w}}_{i,t-1}^\T  \bvltn{\phi}_t - y_t | }{2 (\epsilon_{i,t-1}^u + \nu )}}^\beta
		\end{equation}
		Since the LHS of \eqref{eqn_untitled4} is larger than or equal to $r_j$, to show \eqref{eqn_untitled4}, it suffices to show
		\begin{equation} \label{eqn_untitled5}
		\ r_j
		>
		\ r_i \cdot \max\parenthesesbig{1, \frac{\alpha}{2} \cdot \frac{ \norm{\bvltn{\phi}_t}^{-1} | \wht{\bv{w}}_{i,t-1}^\T  \bvltn{\phi}_t - y_t | }{\epsilon_{i,t-1}^u + \nu }}^\beta
		\end{equation}
		Note that we could have the following
		\begin{equation}
		\begin{split}
		&\frac{ \norm{\bvltn{\phi}_t}^{-1} | \wht{\bv{w}}_{i,t-1}^\T  \bvltn{\phi}_t - y_t | }{ \epsilon_{i,t-1}^u + \nu } \\
		= &\norm{\bvltn{\phi}_t}^{-1} \frac{ | n_t + \bvltn{\epsilon}_{i, t-1}^\T \bvltn{\phi}_t | }{ \epsilon_{i,t-1}^u + \nu } \\
		\leq & \norm{\bvltn{\phi}_t}^{-1} \frac{ | n_t | + \norm{\bvltn{\epsilon}_{i, t-1}} \norm{\bvltn{\phi}_t} }{ \epsilon_{i,t-1}^u + \nu } \\
		= &\frac{ | n_t | }{\norm{\bvltn{\phi}_t} ( \epsilon_{i,t-1}^u + \nu )} + \frac{ \norm{\bvltn{\epsilon}_{i, t-1}}  }{ \epsilon_{i,t-1}^u + \nu } \\
		\leq &\frac{ | n_t | }{\norm{\bvltn{\phi}_t} ( \norm{\bvltn{\epsilon}_{i, t-1}} + \nu )} + 1
		\end{split}
		\end{equation}
		where the first line holds since $y_t = \bv{w}_i^\T \bvltn{\phi}_t + n_t$, and $\bvltn{\epsilon}_{i,t-1} = \bv{w}_i - \wht{\bv{w}}_{i,t-1}$; the last line holds since $\epsilon_{i,t-1}^u > \norm{\bvltn{\epsilon}_{i, t-1}}$ from Theorem \ref{thrm_validUpperbound}. Since we let $\alpha = 2, \beta = 1$, to ensure \eqref{eqn_untitled5} holds, it suffices to ensure the following holds:
		\begin{equation} \label{eqn_untitled7}
		r_j > r_i \parenthesesbig{\frac{ | n_t | }{\norm{\bvltn{\phi}_t} ( \norm{\bvltn{\epsilon}_{i, t-1}} + \nu )} + 1}
		\end{equation}
		Since $r_i = \norm{\bvltn{\phi}_t}^{-1} | y_t - \wht{\bv{w}}_{i, t-1}^\T \bvltn{\phi}_t | ,  r_j = \norm{\bvltn{\phi}_t}^{-1} | y_t - \wht{\bv{w}}_{j, t-1}^\T \bvltn{\phi}_t |$, $\bvltn{\epsilon}_{j, t-1} = \bv{w}_j - \wht{\bv{w}}_{j,t-1}$ and $y_t = \bv{w}_i^\T \bvltn{\phi}_t + n_t$, we have
		\begin{align}
		r_i &= \norm{\bvltn{\phi}_t}^{-1} |n_t + \bvltn{\epsilon}_{i, t-1}^\T \bvltn{\phi}_t |\\
		r_j &= \norm{\bvltn{\phi}_t}^{-1} |n_t + (\bv{w}_i - \bv{w}_j)^\T \bvltn{\phi}_t +  \bvltn{\epsilon}_{j, t-1}^\T \bvltn{\phi}_t |
		\end{align}
		So, \eqref{eqn_untitled7} is equivalent to 
		\begin{equation} \label{eqn_untitled8}
		|n_t + (\bv{w}_i - \bv{w}_j)^\T \bvltn{\phi}_t +  \bvltn{\epsilon}_{j, t-1}^\T \bvltn{\phi}_t |
		> \\
		|n_t + \bvltn{\epsilon}_{i, t-1}^\T \bvltn{\phi}_t | \parenthesesbig{\frac{ | n_t | }{\norm{\bvltn{\phi}_t} ( \norm{\bvltn{\epsilon}_{i, t-1}} + \nu )} + 1}
		\end{equation}
		Note that in \eqref{eqn_untitled8}, we can see
		\begin{equation} \label{eqn_untitle51}
		LHS \geq |(\bv{w}_i - \bv{w}_j)^\T \bvltn{\phi}_t| - |n_t | - |\bvltn{\epsilon}_{j, t-1}^\T \bvltn{\phi}_t |
		\end{equation}
		\begin{equation} \label{eqn_untitle50}
		\begin{split}
		RHS 
		&\leq |n_t| + |\bvltn{\epsilon}_{i, t-1}^\T \bvltn{\phi}_t | +     \frac{ \parentheses{|n_t| + \norm{\bvltn{\epsilon}_{i, t-1}} \norm{\bvltn{\phi}_t} |} | n_t | }{\norm{\bvltn{\phi}_t} ( \norm{\bvltn{\epsilon}_{i, t-1}} + \nu )} \\
		&= |n_t| {+} |\bvltn{\epsilon}_{i, t-1}^\T \bvltn{\phi}_t | {+} \frac{\norm{\bvltn{\epsilon}_{i, t-1}} | n_t | }{\norm{\bvltn{\epsilon}_{i, t-1}} {+} \nu } {+} \frac{ | n_t |^2 }{\norm{\bvltn{\phi}_t} ( \norm{\bvltn{\epsilon}_{i, t-1}} {+} \nu )} \\
		&< |n_t| + |\bvltn{\epsilon}_{i, t-1}^\T \bvltn{\phi}_t | + | n_t | + \frac{ | n_t |^2 }{\norm{\bvltn{\phi}_t} \nu} \\
		&= 2|n_t| + |\bvltn{\epsilon}_{i, t-1}^\T \bvltn{\phi}_t | + \frac{ | n_t |^2 }{\norm{\bvltn{\phi}_t} \nu}
		\end{split}
		\end{equation}
		Considering \eqref{eqn_untitle51} and \eqref{eqn_untitle50}, we can see to ensure \eqref{eqn_untitled8} holds, it suffices to let
		\begin{equation} \label{eqn_untitled9}
		|\bvltn{\epsilon}_{i, t-1}^\T \bvltn{\phi}_t | + |\bvltn{\epsilon}_{j, t-1}^\T \bvltn{\phi}_t | + 3 |n_t| - |(\bv{w}_i - \bv{w}_j)^\T \bvltn{\phi}_t| + \frac{ | n_t |^2 }{\norm{\bvltn{\phi}_t} \nu} \leq 0
		\end{equation}
		From Assumption \ref{asmp_noise}, \ref{asmp_data}, \ref{asmp_NoAmbiguous}, we have $\norm{\bvltn{\phi}_t} \leq \phi_{\max}, |n_t| \leq n_{\max}, |(\bv{w}_i - \bv{w}_j)^\T \bvltn{\phi}_t| \geq \psi, \frac{| n_t |}{\norm{\bvltn{\phi}_t}} \leq \frac{1}{S_{\min}} $. Applying these bounds to \eqref{eqn_untitled9}, we can see to ensure \eqref{eqn_untitled8} holds, it suffices to let
		\begin{equation} \label{eqn_untitled10}
		\parenthesesbig{\norm{\bvltn{\epsilon}_{i, t-1}} + \norm{\bvltn{\epsilon}_{j, t-1}}} \phi_{\max} + 3 n_{\max} - \psi + \frac{n_{\max}}{S_{\min} \nu} \leq 0
		\end{equation}
		So, to ensure \eqref{eqn_untitled10} holds, it suffices to have $\forall i \in [\nsys]$
		\begin{equation} \label{eqn_untitled11}
		\norm{\bvltn{\epsilon}_{i, t-1}} \leq \frac{1}{2 \phi_{\max}} \parenthesesbig{\psi - \frac{n_{\max}}{\nu S_{\min}} - 3n_{\max}} = \epsilon'
		\end{equation}
		Tracing all the way back, we can see when \eqref{eqn_untitled11} holds for $\forall i \in [\nsys]$, \eqref{eqn_untitled3} would hold, therefore we could assign data $\curlybrackets{\bvltn{\phi}_t, y_t}$ generated by subsystem $i$ to candidate $i$, i.e. $\wht{\sysidx}_t = {\sysidx}_t$.
	\end{pf}
	
	\subsection{{Proof for Theorem \ref{thrm_LocalConvergence}}}
	
	\begin{pf}
	For ease of explanation, we let \textbf{Correct Assignment Always (CAA)} be the event of correct assignment at every time step, which is exactly result (i). 
	Note that in the claims of this theorem, result (ii) is a direct consequence of result (i) according to Theorem \ref{thrm_PartialConstant}. To prove this theorem, it suffices to prove \textbf{CAA} happens with probability at least $1 - 2 \nsys \sqrt{ \frac{N_R}{\epsilon'^2} \parenthesesbig{\epsilon_0^2 + \frac{N_R}{S_{\min}^2}} }$ It's difficult to evaluate \textbf{CAA} directly, so we will evaluate \textbf{CAA} only on the \textbf{perfect event trajectory (PET)}: ``at every time step (including 0), all candidates have accurate enough estimate after updates such that we can make correct assignment at next time step according to Lemma \ref{lemma_noMisassignment}; at every time step, we can make correct assignment''. Since \textbf{CAA} occurs whenever \textbf{PET} occurs, a lower bound on $P(\textbf{PET})$ would also be a lower bound on $P(\textbf{CAA})$. We will show that to evaluate $P(\textbf{PET})$, it suffices to study each candidate separately and then combine them altogether. We will illustrate this with a toy example and then generalize it to general cases.

	\begin{table*}[h]
	\caption{Perfect Event Trajectory (\textbf{PET})}
	\label{tbl_PET}
	\centering
	\begin{tabular}{l|lll}
		Time Indices & & Correct Assign  & \quad \quad \  \, Accurate Enough Estimation After Update\\
		\hline
		$(t=0)$: & & NA & \quad \quad \  \, $\norm{\bvltn{\epsilon}_{1,0}}^2, \norm{\bvltn{\epsilon}_{2,0}}^2 \leq \epsilon'^2$ \\
		$(t=1)$: &$ \overset{wp1}{\longrightarrow} $ & $\wht{\sysidx}_1 = 1 $ &
		$\longrightarrow \quad \norm{\bvltn{\epsilon}_{1,1}}^2 , \norm{\bvltn{\epsilon}_{2,1}}^2 \leq \epsilon'^2 \text{ after update of } \wht{\bv{w}}_{1,0}$\\
		&&& \vdots \\
		$(t=t_1-1)$: &$ \overset{wp1}{\longrightarrow} $ & $\wht{\sysidx}_{t_1-1} = 1 $ &
		$\longrightarrow \quad \norm{\bvltn{\epsilon}_{1,t_1-1}}^2 , \norm{\bvltn{\epsilon}_{2,t_1-1}}^2 \leq \epsilon'^2 \text{ after update of } \wht{\bv{w}}_{1,t_1-2}$\\
		$(t=t_1)$: &$ \overset{wp1}{\longrightarrow} $ & $\wht{\sysidx}_{t_1} = 2 $ &
		$\longrightarrow \quad \norm{\bvltn{\epsilon}_{1,t_1}}^2 , \norm{\bvltn{\epsilon}_{2,t_1}}^2 \leq \epsilon'^2 \text{ after update of } \wht{\bv{w}}_{1,t_1-1}$\\
		&&& \vdots \\
		$(t=t_2-1)$: &$ \overset{wp1}{\longrightarrow} $ & $\wht{\sysidx}_{t_2-1} = 2 $ &
		$\longrightarrow \quad \norm{\bvltn{\epsilon}_{1,t_2-1}}^2 , \norm{\bvltn{\epsilon}_{2,t_2-1}}^2 \leq \epsilon'^2 \text{ after update of } \wht{\bv{w}}_{1,t_2-2}$\\
		$(t=t_2)$: &$ \overset{wp1}{\longrightarrow} $ & $\wht{\sysidx}_{t_2} = 1 $&
		$\longrightarrow \quad \norm{\bvltn{\epsilon}_{1,t_2}}^2 , \norm{\bvltn{\epsilon}_{2,t_2}}^2 \leq \epsilon'^2 \text{ after update of } \wht{\bv{w}}_{1,t_2-1}$\\
		&&& \vdots \\							
	\end{tabular}
	\end{table*}
	
	Assume there are only two subsystems 1 and 2 in the hybrid SARX system. Subsystem 1 dominates at time $\curlybrackets{1, 2, \dots, t_1-1, t_2, t_2+1, \dots}$ and subsystem 2 dominates at $\curlybrackets{t_1, t_1+1, \dots, t_2-1}$. This is to say, there is a switching from 1 to 2 at time $t_1$, and 2 back to 1 at time $t_2$. Now, consider the \textbf{PET} in Table \ref{tbl_PET}. In this table, time indices are listed on the left of the vertical separator, and the events occur at different time steps are listed on the right. The ``\emph{Correct Assign}'' column lists the events of making correct assignment at different time steps. The ``\emph{Accurate Enough Estimation After Update}'' column lists the events of accurate enough (below $\epsilon'$) estimation after update. ``$\overset{wp1}{\longrightarrow}$'' means event $\curlybrackets{\norm{\bvltn{\epsilon}_{1,t-1}}^2 , \norm{\bvltn{\epsilon}_{2,t-1}}^2 \leq \epsilon'^2 \text{ after update of } \wht{\bv{w}}_{i,t-2} \text{ for some i}}$ at time $t-1$ will lead to event $\curlybrackets{\wht{\sysidx}_t = \sysidx_t}$, i.e. making correct assignment at time $t$, with probability 1, whose justification is given in Lemma \ref{lemma_noMisassignment}. ``$\longrightarrow$'' means with certain probability, current correct assignment will make estimates accurate enough after update. From Table \ref{tbl_PET}, we can see the randomness in \textbf{PET} only come from all the events in the ``\emph{Accurate Enough Estimation After Update}'' column. In another way, to evaluate $P(\textbf{PET})$, it's equivalent to evaluate the probability that for every time step, after updating estimate with data from correct subsystem, the new estimation error will be smaller than $\epsilon'^2$. To see things more clearly, the events we want to evaluate have the following properties
	\begin{enumerate}
	\setlength{\itemsep}{.1pt}		
	\item At time $t=1,2, \dots, t_1-1$, only $\norm{\bvltn{\epsilon}_{1,t}}^2$ is changing while $\norm{\bvltn{\epsilon}_{2,t}}^2 = \norm{\bvltn{\epsilon}_{2,0}}^2$ is unchanged. And we always have $\norm{\bvltn{\epsilon}_{1,t}}^2, \norm{\bvltn{\epsilon}_{2,t}}^2 \leq \epsilon'^2$
	\item At time $t=t_1, t_1+1, t_2-1$, only $\norm{\bvltn{\epsilon}_{2,t}}^2$ is changing while $\norm{\bvltn{\epsilon}_{1,t}}^2 = \norm{\bvltn{\epsilon}_{1,t_1-1}}^2 $ is unchanged. And we always have $\norm{\bvltn{\epsilon}_{1,t}}^2, \norm{\bvltn{\epsilon}_{2,t}}^2 \leq \epsilon'^2$
	\item At time $t=t_2, t_2+1, \dots$, only $\norm{\bvltn{\epsilon}_{1,t}}^2$ is changing while $\norm{\bvltn{\epsilon}_{2,t}}^2 = \norm{\bvltn{\epsilon}_{2,t_2-1}}^2$ is unchanged. And we always have $\norm{\bvltn{\epsilon}_{1,t}}^2, \norm{\bvltn{\epsilon}_{2,t}}^2 \leq \epsilon'^2$
	\item Additionally, we have $\norm{\bvltn{\epsilon}_{1,0}}^2, \norm{\bvltn{\epsilon}_{2,0}}^2 \leq \epsilon'^2$
	\end{enumerate}
	Now consider the following fictitious Scenario (A): with $\norm{\bvltn{\epsilon}_{1,0}}^2, \norm{\bvltn{\epsilon}_{2,0}}^2 \leq \epsilon'^2$, let $\bvltn{\Phi}_1$ and $\bvltn{\Phi}_2$ denote the data we assigned to candidate 1 $\wht{\bv{w}}_{1, t}$ and candidate 2 $\wht{\bv{w}}_{2, t}$ respectively; then we first update $\wht{\bv{w}}_{1, 0}$ using $\bvltn{\Phi}_1$ and then update $\wht{\bv{w}}_{2, 0}$ with $\bvltn{\Phi}_2$ as if there is always only one subsystem during this course. With the properties listed above, we can see $P(\textbf{PET}) = P(C_1 \bigcap C_2)$ where $C_i$ is the event ``candidate i always have error smaller than $\epsilon'^2$ in Scenario (A)''. Since $C_1$, $C_2$ corresponds to applying algorithm to single subsystem, we can see $P(C_1)$ and $P(C_2)$ can be lower bounded by the probability in Lemma \ref{lemma_ProbBoundforSingleSys}. Therefore, we have
	\begin{equation}
	P(\textbf{CAA}) \geq P(\textbf{PET}) \geq 
	1-P(C_1^c) - P(C_2^c) \geq 1- 2 \cdot 2 \sqrt{ \frac{N_R}{\epsilon'^2} \parenthesesbig{\epsilon_0^2 + \frac{N_R}{S_{\min}^2}} }
	\end{equation}
	
	Now for the more general SARX model with $\nsys$ subsystems, we could generalize the argument above, and end up getting
	\begin{equation}
	P(\textbf{CAA}) \geq P(\textbf{PET}) \geq 1- \nsys \cdot 2 \sqrt{ \frac{N_R}{\epsilon'^2} \parenthesesbig{\epsilon_0^2 + \frac{N_R}{S_{\min}^2}} }
	\end{equation}	
	Finally, with the argument we made at the beginning of the proof, we can see the proof for this theorem is done. 
	\end{pf}
	
	\subsection{{Proof for Corollary \ref{crlry_LocalConvergenceWONoise}}}
	
	\begin{pf}
	When there is no noise, Assumption \ref{asmp_SNRUpperBound} which is the building block to the local convergence result Theorem \ref{thrm_LocalConvergence} is no longer valid as $S_{\min}$ and $S_{\max}$ will both go to $\infty$ and $\kappa_{\max} \geq \frac{S_{\max}}{S_{\min}}$ is no longer well defined. However, in this case, we can prove variants of Lemma \ref{lemma_Supermartingale} and Lemma \ref{lemma_ProbBoundforSingleSys} without relying on Assumption \ref{asmp_SNRUpperBound}.
	
	Specifically, in the proof of Lemma \ref{lemma_Supermartingale}, if $n_t=0$, the last term $\frac{1}{S_{\min}^2}$ in \eqref{eqn_untitle39} would vanish and we proved the supermartingale directly and thus Lemma \ref{lemma_Supermartingale} holds without relying on Assumption \ref{asmp_SNRUpperBound}. And in the proof of Lemma \ref{lemma_ProbBoundforSingleSys}, all the term $\frac{N_R}{S_{\min}^2}$ would vanish due to the absence of noise. So the claim of Lemma \ref{lemma_ProbBoundforSingleSys} would be assume $\norm{\bvltn{\epsilon}_{i,0}} \leq \epsilon_0$ such that $\sqrt{N_R\epsilon_0^2 } \leq \epsilon'$, then for $\forall t$ we have
	\begin{equation}
	P \parenthesesbig{\bigcap_{\tau = 1}^t \curlybracketsbig{\norm{\bvltn{\epsilon}_{i, \tau}}^2 \leq \epsilon'^2} } \geq 1 - 2 \sqrt{ \frac{N_R}{\epsilon'^2} \epsilon_0^2 }
	\end{equation}
	
	We could apply this variant of Lemma \ref{lemma_ProbBoundforSingleSys} to proof for Theorem \ref{thrm_LocalConvergence} directly and get the probability bound $1 - 2 \nsys \sqrt{ \frac{N_R}{\epsilon'^2} \epsilon_0^2 }$. Finally we can get \eqref{eqn_untitle44} in the corollary simply by letting $\sigma_n=0$ in the partial convergence result Theorem \ref{thrm_PartialConstant}.
	\end{pf}

	\section{Proofs for Extension Results Theorem \ref{thrm_IncreaseMonteCarlo}}
	
	\subsection{{Proof for Theorem \ref{thrm_IncreaseMonteCarlo}}}
	
	\begin{pf}
	From the \eqref{eqn_untitle45}, we see $\bvltn{\epsilon}_{t} = \bv{b} - \bv{A} \bv{n}_t$ for some $\bv{A}$ and $\bv{b}$. For some $\tilde{\epsilon}>0$, we have
	\begin{equation}
	\begin{split}
	P(\norm{\bvltn{\epsilon}_{t}} \leq \tilde{\epsilon} ) 
	&= P(\indicator \curlybrackets{\norm{\bv{A} \bv{n}_t - \bv{b}} \leq \tilde{\epsilon} }) \\
	&= \expctn \squarebrackets{\indicator \curlybrackets{\norm{\bv{A} \bv{n}_t - \bv{b}} \leq \tilde{\epsilon} }}
	\end{split}		
	\end{equation}
	With Monte Carlo samples of $\bv{n}_t, \curlybrackets{\bv{n}_t^{(i)}}_{i=1}^{N_t}$, according to Hoeffding's inequality, we have
	\begin{equation}
	\begin{split}
		&P \parenthesesbig{\frac{1}{N_t} \sum_{i=1}^{N} \indicator \curlybrackets{\norm{\bv{A} \bv{n}_t^{(i)} {-} \bv{b}} {\leq} \tilde{\epsilon} } - P(\norm{\bvltn{\epsilon}_{t}} \leq \tilde{\epsilon} ) {\leq} \zeta_1^t} \\
		&\geq 1- \exp \parentheses{-2N_t \zeta_1^{2t}}
	\end{split}
	\end{equation}
	Let $\tilde{\epsilon} = \epsilon_t^u = \max \norm{A \bv{n}_t^{(i)} - \bv{b}}$, and note that $N_t \geq \frac{\zeta_2 t}{2\zeta_1^{2t}}$, we have
	\begin{equation} 
		P \parenthesesbig{ P(\norm{\bvltn{\epsilon}_{t}} \leq \tilde{\epsilon} ) \geq 1 - \zeta_1^t}
		\geq 1- \exp \parentheses{-\zeta_2 t}
	\end{equation}
	Using union bound, we have
	\begin{equation} \label{eqn_untitle47}
		P \parenthesesbig{\bigcap_{t=1}^\infty \curlybracketsbig{ P(\norm{\bvltn{\epsilon}_{t}} \leq \tilde{\epsilon} ) \geq 1 - \zeta_1^t} }
		\geq 1- \frac{\exp \parentheses{-\zeta_2}}{1- \exp \parentheses{-\zeta_2}}
	\end{equation}
	For the event inside $P(\cdot)$, according to union bound, we have
	\begin{equation} \label{eqn_untitle46}
		\bigcap_{t=1}^\infty \curlybracketsbig{ P(\norm{\bvltn{\epsilon}_{t}} \leq \tilde{\epsilon} ) \geq 1 - \zeta_1^t} 
		\Rightarrow \\
		P \parenthesesbig{ \bigcap_{t=1}^\infty \curlybracketsbig{ \norm{\bvltn{\epsilon}_{t}} \leq \tilde{\epsilon} } \geq 1 - \frac{\zeta_1}{1-\zeta_1}} 
	\end{equation}
	Therefore, plugging \eqref{eqn_untitle46} into \eqref{eqn_untitle47}, we could get \eqref{eqn_untitle48}
	\end{pf}

\end{document}